\documentclass[11pt,a4paper]{article}
\usepackage{amsmath}
\usepackage{amsfonts}
\usepackage{amssymb}
\newcommand{\nn}{\nonumber}
\newcommand{\no}{\nonumber\\}
\newcommand{\be}{\begin{equation}}
\newcommand{\ee}{\end{equation}}
\newcommand{\ba}{\begin{eqnarray}}
\newcommand{\ea}{\end{eqnarray}}

\def\gl#1{(\ref{#1})}

\def\i{{\rm i}}
\def\d{{\rm d}}

\def\reg{{\rm reg}\,}

\def\di{\not\!\partial}
\def\A{\not\!\!A}
\def\b{\not\!b}
\def\p{\not\!p}

\def\ve{\varepsilon}

\def\ie{{\rm i.e.}\ }
\def\eg{{\rm e.g.}\ }
\def\tr{{\rm tr}\,}
\def\p{p\!\!/}

\def\x{\;\!}
\def\bfpi{\mbox{\boldmath $\pi$}}
\date{}
\begin{document}
\begin{center}
{\Large\bf Bare and Induced Lorentz \& CPT Invariance Violations in QED}
\end{center}
\bigskip

\centerline{J. ALFARO$^{\,1}$,  A.A. ANDRIANOV$^{\,2}$,
 M. CAMBIASO$^{\,3}$, P. GIACCONI$^{\,4}$, R. SOLDATI$^{\,4}$}
\smallskip
\centerline{$^{1}$ {\it Facultad de Fisica, Pontificia Universidad Cat\'olica de Chile,}}
\centerline{\it Vicu\~{n}a Mackenna 4860 Macul, Santiago de Chile, Casilla 306, Chile}
\smallskip
\centerline{$^{2}$ {\it V.A. Fock Department of Theoretical Physics, Sankt-Petersburg State University,}}
\centerline{\it 198504 Sankt-Petersburg, Russia}
\smallskip
\centerline{\it Departament d'Estructura i Constituents de la Materia,}
\centerline{\it  Universitat de Barcelona, 08028, Barcelona, Spain}
\centerline{$^{3}$ {\it Instituto de Ciencias Nucleares, Universidad Nacional Aut\'onoma de M\'exico,}}
\centerline{\it A. Postal 70-543, M\'exico D.F., M\'exico}
\smallskip
\centerline{\it Departamento de Ciencias Fisicas,}
\centerline{\it Universidad Andr\'es Bello, Av. Rep\'ublica 252, Santiago, Chile}
\smallskip
\centerline{$^{4}$ {\it Dipartimento di Fisica, Universit\'a di Bologna}}
\centerline{\it Istituto Nazionale di Fisica Nucleare, Sezione di Bologna,}
\centerline{\it 40126 Bologna, Italia}

\vskip .5cm

\abstract
We consider QED in a constant axial vector background (\AE ther). Further Lorentz invariance violations
(LIV) might occur owing to radiative corrections. The phenomenology of this model is
studied, clarifying issues related to the various regularizations employed, with a particular
emphasis on the induced photon mass.
To this concern, it is shown that in the presence of LIV dimensional regularization
may produce a radiatively induced finite photon mass.
The possible physical role of the large momentum  cutoff
is elucidated and the finite temperature radiative corrections
are evaluated. Finally, various
experimental bounds on the parameters of the model are discussed.

\vskip .5cm

\section{{Introduction and Motivations}}

The bounds for the validity of fundamental laws in Physics has attracted more and more
attention and interest in view of remarkable improvements in the experimental technique,
both in laboratory research and in astrophysics \cite{Shore:2004sh}--\cite{ Stecker:2009hj}.
Among others, some important investigations concern possible violations of spacetime symmetries
in vacuum due to the presence of constant backgrounds named \AE thers \cite{Carroll:2009em,Carroll:2009en}.
In practice, the { Lorentz and CPT Invariance Violation} (LIV for short)
in Quantum Electrodynamics \cite{Carroll:1989vb}--
\cite{Myers:2003fd} has not yet been detected \cite{Kostelecky:2002hh}--\cite{Cheng:2006us}.
Nonetheless, it is not even excluded and it might occur \cite{Shore:2004sh}-- \cite{Stecker:2009hj}, \cite{Froggatt:1991ft}--\cite{Urrutia:2005ty}:
in particular, spontaneous symmetry breaking
\cite{Heisenberg:1957}--\cite{Eguchi:1976iz} may cause LIV after condensation of massless axion-like fields \cite{Andrianov:1994qv}--\cite{ArkaniHamed:2004ar}
and/or of certain vector fields \cite{Will:1972zz}--\cite{Graesser:2005bg} (maybe, of gravitational origin \cite{Kostelecky:2003fs,Mukhopadhyay:2005gb}),
as well as short distance spacetime asymmetries may come from string \cite{Kostelecky:1988zi}--\cite{Bertolami:1998dn}
and quantum gravity effects \cite{AmelinoCamelia:1996pj}--\cite{Alfaro:2004aa} and non-commutative structure of the spacetime \cite{Carmona:2002iv}--\cite{Das:2005uq}.
\subsection{Lorentz and Gauge Invariance Violations}
In this paper we shall investigate how a constant axial vector $b_\mu\,,$ a torsion-like
background \AE ther, coupled to massive fermions may change electrodynamics,
due to the vacuum polarization effects.
In the last ten years this issue has been treated
quite extensively in the literature by means of different regularizations
to deal with 1-loop ultraviolet divergences,
with a rich variety of answers
\cite{Jackiw:1999yp}--cite{Brito:2008ec}
following the effective field theory approach,
to the lowest orders in the LIV parameters \cite{Colladay:1996iz,Colladay:1998fq,Kostelecky:2008ts}. This means that, in fact,
within this  approach it is admittedly impossible to predict the actual values
of the LIV structural constants, including their radiatively induced parts,
leaving them to be eventually obtained only from experiments.
In other words, it is just the empirical and phenomenological nature
of latter constants that does explain why their determination really lies beyond any formal argument related to
renormalization of divergent integrals in perturbation theory.
For example, the spatial component of the Chern-Simons four vector $\eta_\mu\,,$ which might constitute another
constant \AE ther background \cite{Carroll:1989vb}, is essentially set equal to zero by the
absence of vacuum birefringence from distant radiogalaxies \cite{Wardle:1997gu}.
In the same manner, the spatial component of the axial vector constant \AE ther $b_\mu$
is severely constrained by the torsion pendulum experiments with polarized electrons
\cite{Heckel:2006ww} to be smaller than the benchmark value $m_e^2/M_{\rm Planck}=2\times10^{-17}$ eV.
Hence the only left narrow possibility is a temporal constant \AE ther, as described by
the axial vector $b_\mu=(b,0,0,0)$ and the Chern-Simons vector $\eta_\mu=(\eta,0,0,0)$ --
the metric tensor is $g_{\mu\nu}={\rm diag}\,(+,-,-,-)\,.$

Of course, since the Chern-Simons lagrangian and the photon mass terms are
local functional of the gauge vector potential,
one possibility \cite{Coleman:1997xq, Coleman:1998en,
Coleman:1998ti,Bonneau:2006ma,Bonneau:2005qg} is to set the  Chern-Simons
four vector $\eta_\mu$
as well as the photon mass $m_\gamma$ exactly equal to zero by assumption, i.e.
as a renormalization prescription, in such a manner to enforce the strong gauge
invariance principle and the Ward identities, even in the occurrence of some
possible Lorentz invariance violating effects. In such a circumstance, perhaps
the simplest and most conservative one, the quantum photodynamics is governed by the usual
Lorentz covariant and gauge invariant Maxwell lagrangian, supplemented by the
gauge fixing terms. In so doing, however,
any possible LIV effect is necessarily confined
inside the spinor matter and does not propagate,
since the photodynamics with the minimal coupling interaction
is dictated by the Lorentz and gauge invariances.

The situation drastically changes if we consider the Maxwell-Chern-Simons modifications
of QED, as originally suggested by Carroll, Field and Jackiw in their seminal paper
\cite{Carroll:1989vb}. In fact, all these modifications of QED are weakly gauge invariant,
in the sense that only the action, but not the lagrangian, is invariant under
the local gauge transformations.
Moreover, the Maxwell-Chern-Simons free radiation field with a temporal
Chern-Simons vector and a massless photon is inconsistent \cite{Andrianov:1994qv,Andrianov:1998wj,Andrianov:1998ay,Adam:2001ma}
since it exhibits acausal and tachyonic
behaviour of the free photons.
Thus, the only way to restore consistency is by means of a tiny but non-vanishing photon mass
$m_{\gamma}\propto b\,.$
In such a case, the strong experimental bounds on the photon mass \cite{Amsler:2008zzb}
definitely endorse a rather stringent limit also on a temporal \AE ther background.
In the recent literature \cite{Altschul:2004gs,Altschul:2005vc,Gabadadze:2004iv,Dvali:2005nt,Altschul:2007kr,Ferrero:2009jb}
a class of Lorentz invariance violating models with photon speed less than one
has been considered and their possible phenomenological consequences have been discussed.
These are models which break gauge invariance.
Specifically, in this paper we show that even if one starts with a gauge invariant
Maxwell lagrangian minimally coupled with LIV fermions, then the 1-loop corrections
might indeed radiatively generate a Chern-Simons term and a Proca mass term for photons.
This entails that the 1-loop effective field theory is safely causal but no
longer gauge invariant.
In other words, a Lorentz invariance violation in the fermion sector
may induce the breaking of gauge invariance of the photon sector.

There are several reasons why this latter possibility, in spite of appearing
somewhat unlikely and unorthodox at a first sight, is actually much more natural and pregnant,
once the Lorentz covariant framework is supposed to be abandoned because
of some still unknown quantum effect occurring at very high energies
and momenta. As a matter of fact,
on the one hand, the Lorentz covariant quantization of the electromagnetic
vector potential must be gauge invariant, in order to fulfill the principles
of the special relativity. On the other hand, the covariant quantization of the abelian
Proca massive vector field is also perfectly consistent and safe, when coupled to a
conserved electric current.
Nevertheless, the photon vector field must be massless, \ie gauge invariant, if the Einstein
postulates with an inertial frame independent unit light velocity have to be filled.
Thus, special relativity and gauge invariance
are tightly related. On the contrary, if the axioms of the special relativity
are somewhere abandoned, as we shall deal here with, then the LIV quantization
of the electromagnetic field quite naturally drives towards a
massive photon in order to guarantee causality and consistency,
as we will  discuss in the present
paper. As a consequence, the very stringent limit on the photon mass
will provide a rather severe bound upon the allowed LIV parameters.

Actually, in order to derive unambiguously any prediction in the LIV models,
one has to put into the game some physically justified
ideas about the high energy behaviour.
The key point of our approach
is that the presence of a constant background \AE ther
modifies the dispersion relations between energy and
momentum, not only at low energies but even more at very high energies and momenta.
This is because of a new phenomenon which is forbidden in the Lorentz covariant framework.
Namely, high energy
electrons, positrons and photons with certain polarizations become acausal, that means
they will never be detected, so that they cannot appear as asymptotic states,
if their momenta exceed the value $\Lambda_e\,\sim\,\frac12\,m_e^2/b\,.$
Moreover, owing to the \AE ther background, high energy  electrons and positrons
with the other polarizations undergo brem\ss trahlung in vacuum, while photons with the other polarization
undergo electron positron pairs creation.
Thus, the very high energy momentum electrons, positrons and photons
are not allowed to appear as physical, stable asymptotic states, on their mass shells.
This special issue has been recently proposed \cite{Andrianov:2009tp} to provide a possible interpretation of the
ATIC, PAMELA, HESS and Fermi data.

This feature suggests that the LIV modification of QED involving
the constant background \AE thers might be trusted at most
as an effective field theory, in spite of
being renormalizable, valid up to
a very high energy momentum scale, beyond which some new and more fundamental Physics will enter into the game.
Since a conservative value for this very high energy momentum scale $\Lambda_e$ will be found to be of the order
$m_e^2/2b\sim10^{\,26}$ eV, it is plausible that the would be new Physics will perhaps concern quantum gravity,
non-commutative field theories
and string theories, as originally suggested \cite{Kostelecky:1988zi,Kostelecky:1994rn,AmelinoCamelia:1996pj} long ago in the literature.
Accordingly, a new fundamental length $\ell_e\,\sim\,\Lambda_e^{-1}$ appears below which
the principles of special relativity could be modified by some new physical phenomenon.
\subsection{LIV Radiative Corrections}
The unavoidable existence of a large momentum cutoff $\Lambda_f\sim\,\frac12\,m_f^2/b$ for the free Dirac
field $\psi_f$ of mass $m_f$ in a constant \AE ther background was discovered by
Kosteleck\'y and Lenhert \cite{Kostelecky:2000mm} and its role was further investigated by other authors
\cite{Andrianov:2001zj,Ebert:2004pq,Lehnert:2003ue,Baeta Scarpelli:2006kd}.
Its very existence, which is of a kinematical and non-analytic nature,
will also provide an essential tool
to understand the meaning and the role of the radiative corrections.
In this paper we shall consider the fermion loops
involving  electron positron virtual pairs and giving rise to
ultraviolet divergent Feynman integrals. Actually, at variance with the covariant case,
their regularization is a nontrivial task,
because there are indeed several inequivalent ways to implement the well known
dimensional \cite{Andrianov:2001zj} and Pauli-Villars \cite{Altschul:2004gs,Altschul:2005vc} regularizations in a LIV divergent Feynman integral.
In particular, dimensional regularization
with the 't Hooft-Veltman-Breitenlohner-Maison recipe to define the algebra of the
$\gamma_5$ matrix in $2\omega$ complex spacetime dimensions
\cite{'tHooft:1972fi,Breitenlohner:1977hr,Breitenlohner:1975hg} does not automatically guarantee
gauge invariance of radiative corrections in a LIV model, as we shall see,
in a manifest contrast to the Lorentz covariant case.

The necessary existence of the very high energy momentum physical bound $\Lambda_e$ in the LIV models
does indeed provide a natural way to understand {\em \` a la} Wilson \cite{Wilson:1973jj}
the integration over the loops momenta. As a matter of fact, the domain $|{\bf p}|<\Lambda_e$
actually corresponds to the integration over physical modes, \ie the large distance Physics,
while the integration over the
outer complementary region in momentum space  will mean the effective inclusion
of the unknown short distance Physics not directly accessible within the LIV model.
In so doing, a divergent photon mass term arises, the finite part of which is arbitrary.
Conversely, in the dimensional regularization method the above physical separation
of the momentum space integration is not at all evident. Nonetheless, owing to the
non-triviality of the $\gamma_5$ algebra in complex spacetime dimensions,
a finite induced Proca mass term for photon just appears from loop integration.

To this concern, it is very important to recall the close relationship between
the \AE ther residual symmetry group and the induced Chern-Simons vector
in the dimensional regularization. Actually, as thoroughly discussed
in ref.~\cite{Andrianov:2001zj}, there are two ways to implement dimensional
regularization in the presence of a temporal \AE ther $b_\mu=(b,0,0,0)$
in the fermionic sector.
The first one, called $\overline{DR}$, leads to a non-vanishing
Chern-Simons vector, is compatible with the existence of the physical
large momentum cutoff, and leaves the residual symmetry group
$O(2\omega-1)$ as the maximal invariance group in momentum space.
The other one, which has been denoted by $\widehat{DR}$,
leads conversely to a null Chern-Simons vector, i.e. to the strong gauge invariance,
but at the price of a further violation of the Lorentz symmetry in the
unphysical dimensions. In this case, the residual symmetry group
is $O(3)\times O(2\omega-3)$ and the existence of the physical
large momentum cutoff is completely disregarded.

Hence, once Lorentz invariance is broken, the enforcement of the strong gauge invariance
with $\eta_\mu=m_\gamma=0$
appears to be somewhat unnatural already at the 1-loop approximation.
As nicely shown in \cite{Andrianov:2001zj}
it looks like a trick {\em ad hoc} involving a double
violation of the Lorentz invariance even in the unphysical spacetime.
Thus, contrary to the first sight impression,
a comparison between the two methods drives to a
finite and non-vanishing result for both the
radiatively generated Chern-Simons vector and the photon mass term, as the most natural
and interesting option, at least in our opinion.
These theoretically estimated finite values can be used
to set a bound on the temporal vector \AE ther $b_\mu$
from the experimental data, notably from the limits on the photon mass.

It is worthwhile to stress again that the
very high momentum region $|{\bf p}|> \Lambda_e\sim m_e^2/2b$
just corresponds to the physically inaccessible domain for the minimal
LIV modification of quantum electrodynamics.
Thus the main results of our analysis is that, once the Lorentz invariance
is broken, both a Chern-Simons term and a photon mass are possible and natural,
as further endorsed by the calculation of the finite temperature 1-loop effective action.
\section{Consistency of Lorentz Invariance Violation}
Before the calculation of the induced
Lorentz invariance violations, let us examine the consistency of LIV
in quantum electrodynamics based on stability of its constituents --
electrons, positrons and photons. To the lowest order it means that,
within the range of validity of Lorentz covariant quantum electrodynamics,
electrons should not undergo a brem\ss trahlung in vacuum $e^- \rightarrow e^-\,\gamma\,,$
\ie photon emission in the absence of external fields,
and real photons should not annihilate in vacuum into real pairs $\gamma \rightarrow e^-\,e^+\,.$
Of course, the latter processes are forbidden in covariant QED but might be allowed if Lorentz
covariance is broken, that is, when a constant \AE ther  background is there.
We shall call the above processes vacuum decays of the fundamental particles,
or simply decays, and the fundamental decaying particles will be denoted by
$\widetilde e^{\,-}\,,\,\widetilde e^{\,+}\,,\,\widetilde\gamma\,.$
\subsection{LIV Effective Lagrangian}
Here we will assume that the leading LIV effects are dominated by the softest interactions
with an \AE ther background coupled to the CPT odd operators of canonical mass dimension
equal to three. Hence we can write in the most general Lagrange density
\ba
{\cal L} = {\cal L}_{{\rm INV}} + {\cal L}_{{\rm LIV}} + {\cal L}_{{\rm AS}}
\label{lagrangian}\\
{\cal L}_{{\rm INV}} = -\,{\textstyle\frac14}\,F^{\alpha\beta}(x)F_{\alpha\beta}(x)
+ {\textstyle\frac12}\,m^2_\gamma\,A_\nu(x)A^\nu(x)
+ \bar\psi(x)[\,\i\!\!\di + e\A(x)-m_e\,]\,\psi(x)\\
{\cal L}_{{\rm LIV}} =
{\textstyle\frac12}\,\eta_\alpha A_\beta(x)\tilde F^{\,\alpha\beta}(x)
+ b_\mu\,\bar\psi(x)\,\gamma_5\,\gamma^{\,\mu}\,\psi(x)\\
{\cal L}_{{\rm AS}} = A^\mu(x)\,\partial_\mu B(x) +
{\textstyle\frac12}\,\varkappa\,B^2(x)
\ea
where $A_\mu$ and $\psi(x)$ stand for the LIV photon and electron positron fields respectively,
$e>0$ is the positron charge, $\tilde F^{\,\alpha\beta}(x)={\textstyle\frac12}\,
\varepsilon^{\,\alpha\beta\rho\sigma}\,F_{\,\rho\sigma}(x)$ is the dual field strength,
while $B$ is the auxiliary St\"uckelberg scalar field. Notice that we have included
the Proca mass term for the photon in the Lorentz invariant lagrangian ${\cal L}_{{\rm INV}}$
because, as we shall
discuss in the sequel, the latter is required by self-consistency, i.e. stability and causality,
of such a LIV extension of QED. Moreover, as we shall see, it is generally radiatively induced
by the LIV spinor lagrangian
$b_\mu\,\bar\psi(x)\,\gamma_5\,\gamma^{\,\mu}\,\psi(x)\,.$
However, we will not elaborate here its dynamical origin: depending on whether it is generated
by a Higgs mechanism or it is a fundamental Proca mass, quite different experimental bounds
\cite{Amsler:2008zzb} can be applied to it
\footnote{We also leave the room for the photon mass generation in a plasma like medium.}.
The auxiliary St\"uckelberg lagrangian ${\cal L}_{{\rm AS}}\,,$
which further violates gauge invariance beyond the photon mass term,
has been necessarily introduced to provide, just owing to the so called St\"uckelberg trick,
the simultaneous occurrences of power counting renormalizability and perturbative unitarity.
\footnote{
For an equivalent BRST treatment see for example \cite{Collins:1984xc}.}

Hereafter, the LIV constant vectors $\eta_\mu$ and $b_\mu$ -- the \AE ther --
are supposed to be universal and proportional $\eta_\mu \propto b_\mu$, including both the classical background
$\eta^{\,c\ell}_\mu, b_\mu^{\,c\ell}$ as well as the QED radiative corrections
$\eta_\mu = \eta^{\,c\ell}_\mu + \Delta \eta_\mu,\ b_\mu = b_\mu^{\,c\ell} +\Delta b_\mu\,.$

As a matter of fact, if LIV manifests itself as a fundamental phenomenon in the large scale Universe,
or it is a result of condensation of axial vector/axion gradient fields,
it is quite plausible that  LIV is induced universally by different species
of fermion fields coupled to the very same axial vector $b^{\,\mu}\,,$
albeit with different magnitudes depending upon flavors.
Then both LIV vectors become \cite{Jackiw:1999yp,PerezVictoria:1999uh,PerezVictoria:2001ej,Volovik:1999up,
Chaichian:2000eh,Chung:2001mb,BaetaScarpelli:2001ix,Bonneau:2000ai,Klinkhamer:2004hg,Andrianov:2001zj} collinear, \ie $\eta_{\mu}\propto b_{\mu}\,.$
Meantime it has been found \cite{Andrianov:1994qv,Andrianov:1998wj,Andrianov:1998ay,Andrianov:2001zj,Adam:2001ma,Kostelecky:2000mm} that a consistent quantization
of photons just requires the CS vector to be spacelike,
whereas for the consistency of the spinor free field theory a spacelike
axial vector $b^{\,\mu}$ is generally not allowed, but for the pure spacelike case which,
however, is severely bounded by the experimental data \cite{Dehmelt:1999jh,Phillips:2000dr}.

As already mentioned, to generally provide the self-consistency of the present LIV model,
one has to introduce \cite{Alfaro:2006dd} in the above lagrangian
(\ref{lagrangian}) also a photon mass, for assembling both a bare and an induced masses
$m^2_\gamma=m^2_0 + \Delta m^2_\gamma\,.$
It happens (see below) that the induced photon mass squared $\Delta m^2_\gamma$
is $O(\alpha b^2)$  where $\alpha = e^2/4\pi$ is the fine structure constant.
Thus, it turns out that a minimal breaking of the Lorentz and CPT symmetries
in the spinor matter sector leads to the loss of gauge invariance in the photon sector.

As we shall better see below, all the above mentioned new features will eventually result
in the Lorentz covariance violating modifications
of the classical equations for particle mass shells, \ie
LIV modified transverse photons ($\,\widetilde\gamma\,$)
\be
(k^2 - m^2_\gamma )^2 - (\eta\cdot k)^2 + \eta^2 k^2 = 0\ ,
\label{eqphot}
\ee
and LIV modified electrons and positrons ($\,\widetilde e^{\;\mp}\,$)
\be
(p^2 + b^2 -m^2_e)^2 - 4(b\cdot p)^2 + 4b^2 m_e^2 = 0\ .
\label{eqel}
\ee
Let us examine the possibilities:
$(i)$\quad $\widetilde e^{\,\pm}\ \to\ \widetilde e^{\,\pm}\,\widetilde\gamma\,,$
namely the triggered  decay
\footnote{It can be qualified as either a Cherenkov type radiation
from superluminal $\widetilde e^{\,\pm}\,,$ if LIV is dominating for fermions,
or a brem\ss trahlung of photons in an axial vector background, when fermions are not superluminal.
In  order to avoid any semantic ambiguity, we shall dub it  {LIV decay}.}
of LIV modified electrons moving with emission of modified photons
$\widetilde\gamma\,.$ In the Lorentz covariant QED it would correspond to the electron decay
in its rest frame, which is impossible due to energy conservation. As well, one can also search for:
$(ii)$\quad $\widetilde\gamma\ \to\ \widetilde e^{\,+} \widetilde e^{\,-}$
{photon decay with pair creation}, a process which is  forbidden for massless photons in conventional QED.
Conversely, it turns out that both kinds of processes may occur in the
Lorentz invariance violating quantum electrodynamics for high energy particles,
as the physics is different in differently moving frames.

To be comprehensible in our analytical calculations and for the sake of description
of the possible phenomenological consequences, we definitely assume the privileged class of rotation invariant
LIV isotropic inertial frames -- the \AE ther -- to be concordant with the rest frame of
Cosmic Microwave Background Radiation (CMBR). Some stability issues
for spacelike anisotropic \AE thers were insofar examined in \cite{Shore:2004sh,Jacobson:2005bg,Kostelecky:2008ts,Bietenholz:2008ni,Stecker:2009hj,
Andrianov:2001zj,Adam:2001ma,Kostelecky:2000mm,Kahniashvili:2008va}.
\subsection{LIV Free Fields Quantization}

The canonical quantization of the LIV spinor field has been thoroughly analyzed
in refs. \cite{Kostelecky:2000mm,Andrianov:2001zj}. Here, for the sake of clarity and completeness, we aim to
shortly develop the canonical quantization of a free massive Proca vector field
with additional LIV Chern-Simons term: namely, a rather new issue that will be called
the canonical quantization of the Proca-Chern-Simons vector field.
The classical Euler-Lagrange field equations that follows from the general
lagrangian (\ref{lagrangian}) take the form
\ba
&&\partial_\lambda F^{\,\lambda\nu}
+\ m^2_\gamma\,A^\nu + \eta_\alpha\widetilde F^{\,\alpha\nu} +
\partial^{\,\nu} B + e\,\bar\psi\,\gamma^{\x\nu}\,\psi\;=\;0
\label{Euler_Lagrange}\\
&&\{\gamma^{\,\mu}[\,{\rm i}\partial_{\,\mu}-b_\mu\,\gamma_5+eA_{\,\mu}(x)\,]-m_e\}\,\psi(x)=0\\
&&\partial_\nu A^\nu\;=\;\varkappa\,B
\ea
After contraction of eq.~(\ref{Euler_Lagrange}) with $\partial_\nu$ we find
\begin{equation}
\left(\x\Box+\varkappa\,m_{\,\gamma}^{\x2}\x\right)B(x)=0
\label{aux}
\end{equation}
whence it follows that the auxiliary St\"uckelberg field is always a decoupled
unphysical real scalar field, which is never involved in the electromagnetic interactions
$\forall\,\varkappa\in\mathbb R$.
Turning now to free field theory, i.e. $e=0\,,$ and momentum space
$$
A^\nu(x)=\int\frac{{\rm d}^4k}{(2\pi)^{3/2}}\,\tilde A^\nu(k)\,{\rm e}^{\,-{\rm i} k\cdot x}\qquad
B(x)=\int\frac{{\rm d}^4k}{(2\pi)^{3/2}}\,\tilde B(k)\,{\rm e}^{\,-{\rm i}k\cdot x}
$$
we obtain the momentum space free field equations for the LIV massive vector field
and the auxiliary scalar field
\ba
\left\{g^{\,\lambda\nu}\left(\,k^2-m^2_\gamma\,\right) - k^{\,\lambda}k^{\,\nu} +
{\rm i}\,\varepsilon^{\,\lambda\nu\alpha\beta}\,\eta_\alpha\,k_\beta\right\}
\tilde A_\lambda(k) + {\rm i}\,k^{\,\nu}\,\tilde B(k)=0\\
{\rm i}\,k^{\,\lambda}\,\tilde A_\lambda(k) + \varkappa\,\tilde B(k)=0
\ea
Contractions with $k_{\,\nu}$ and $\eta_{\,\nu}$ respectively yields
\ba
-\,m^2_\gamma\;k\cdot
\tilde A(k) + {\rm i}\,k^{\,2}\,\tilde B(k)=0\\
k\cdot\tilde A(k) = {\rm i}\x\varkappa\,\tilde B(k)\\
\eta\cdot\tilde A(k)\left(\x k^2-m^2_\gamma\x\right) -
(\eta\cdot k)\,k\cdot\tilde A(k)(1-\varkappa^{-1})=0
\ea
Hence, for the specially suitable choice $\varkappa=1$
that greatly simplifies the whole treatment
\footnote{In the covariant canonical quantization of the massless
gauge field this choice corresponds to the well known Feynman gauge,
in which the equations of motion as well as the photon propagator take the simplest form.}
we eventually get
\ba
\tilde B(k)+{\rm i}\,k\cdot\tilde A(k)=0
\label{eqmofgauge1}\\
(\x k^2-m^2_\gamma\x)\;\tilde B(k)=0=
(\x k^2-m^2_\gamma\x)\;\eta\cdot\tilde A(k)
\label{eqmofgauge2}\\
\left\{g^{\,\lambda\nu}\left(\,k^2-m^2_\gamma\,\right) +
{\rm i}\,\varepsilon^{\,\lambda\nu\alpha\beta}\,\eta_\alpha\,k_\beta\right\}
\tilde A_\lambda(k)=0
\label{eqmofgauge}
\ea
Consider now the kinetic $4\times4$ hermitian matrix $\mathbb K$ with
matrix elements
\ba
K_{\,\lambda\nu}\equiv
g_{\,\lambda\nu}\left(\,k^2-m^2_\gamma\,\right) +
{\rm i}\,\varepsilon_{\,\lambda\nu\alpha\beta}\,\eta^\alpha\,k^{\x\beta}
\ea
which satisfies
\ba
K_{\,\lambda\nu}=K^{\,\ast}_{\,\nu\lambda}
\ea
The Levi-Civita symbol in the four dimensional Minkowski spacetime is normalized according to
\begin{equation}
\varepsilon_{\,0123} = -\,\varepsilon^{\,0123}\equiv 1
\nn
\end{equation}
in such a way that
\medskip
\begin{eqnarray}
\varepsilon^{\,\mu\nu\alpha\beta}\,\varepsilon_\mu^{\ \lambda\rho\sigma}&=&
-\,g^{\,\nu\lambda}\,g^{\,\alpha\rho}\,g^{\,\beta\sigma} -
g^{\,\alpha\lambda}\,g^{\,\beta\rho}\,g^{\,\nu\sigma} -
g^{\,\beta\lambda}\,g^{\,\nu\rho}\,g^{\,\alpha\sigma}\no
&+& g^{\,\nu\rho}\,g^{\,\alpha\lambda}\,g^{\,\beta\sigma} +
g^{\,\alpha\rho}\,g^{\,\beta\lambda}\,g^{\,\nu\sigma} +
g^{\,\beta\rho}\,g^{\,\nu\lambda}\,g^{\,\alpha\sigma}
\nn
\end{eqnarray}
Now, it turns out that we have
\ba
&&S^{\,\nu}_{\;\lambda}\ \equiv\ \varepsilon^{\,\mu\nu\alpha\beta}\,\eta_{\x\alpha}\,k_{\x\beta}\,
\varepsilon_{\,\mu\lambda\rho\sigma}\,\eta^{\x\rho}\,k^{\x\sigma}\ =\\
&& \delta^{\,\nu}_{\;\lambda}\,{\mbox{\tt D}}
+ k^{\,\nu}\,k_{\x\lambda}\,\eta^2 + \eta^{\,\nu}\,\eta_{\x\lambda}\,k^2
- \eta\cdot k\,(\x\eta_{\,\lambda}\,k^{\x\nu} + \eta^{\x\nu}\,k_{\,\lambda}\x)
\nn
\ea
where
$$
{\mbox{\tt D}}\;\equiv\;(\x\eta\cdot k)^2-\eta^2\,k^2\;=\;\textstyle\frac12\;S^{\,\nu}_{\;\;\nu}
$$
in such a manner that we find
\ba
S^{\,\nu}_{\;\lambda}\,\eta^{\x\lambda}=S^{\,\nu}_{\;\lambda}\,k^{\x\lambda}\;=\;0\qquad\quad
S^{\,\mu\nu}\,S_{\,\nu\lambda}\;=\;{\mbox{\tt D}}\,S^{\,\mu}_{\;\;\lambda}\qquad\quad
S^{\,\nu}_{\;\;\nu}\;=\;2\,{\mbox{\tt D}}\nn
\ea
while
\ba
S^{\,\mu\lambda}\,\varepsilon_{\,\lambda\nu\alpha\beta}\,\eta^\alpha\,k^{\x\beta}\;
=\;{\mbox{\tt D}}\;\varepsilon^{\,\mu}_{\;\;\nu\alpha\beta}\,\eta^\alpha\,k^{\x\beta}
\ea
Notice that for a timelike and spatial isotropic Axion \AE ther $\eta^{\x\mu}=(\eta,0,0,0)$
we find
$
{\mbox{\tt D}}\;=\;\eta^2\,{\bf k}^2>0\,.
$
Then, to our purpose, it is convenient to introduce the
two orthonormal one dimensional hermitian projectors

\ba
\bfpi^{\,\mu\nu}_{\,\pm}\;\equiv\;
\frac{S^{\,\mu\nu}}{2\,{\mbox{\tt D}}}\;
\pm\;\frac{{\rm i}}{2\sqrt{\mbox{\tt D}}}\,
\varepsilon^{\,\mu\nu\alpha\beta}\,\eta_{\x\alpha}\,k_{\x\beta}\;
=\;\left(\bfpi^{\,\nu\mu}_{\,\pm}\right)^\ast
\ea
which enjoy the properties $\forall\,k^{\,\mu}=(\x k_0,{\bf k}\x)$
\ba
\bfpi^{\,\mu\nu}_{\,\pm}\;\eta_{\x\nu}=\bfpi^{\,\mu\nu}_{\,\pm}\;k_{\x\nu}=0
\qquad\quad g_{\,\mu\nu}\,\bfpi^{\,\mu\nu}_{\,\pm}\;=\;1
\label{projectors1}\\
\bfpi^{\,\mu\lambda}_{\,\pm}\,\bfpi_{\,\pm\,\lambda\nu}\ =\
\bfpi^{\,\mu}_{\,\pm\,\nu}\qquad\quad
\bfpi^{\,\mu\lambda}_{\,\pm}\,\bfpi_{\,\mp\,\lambda\nu}\ =\ 0
\label{projectors2}
\ea
$$
{\mbox{\tt D}}\,\left(\,\bfpi^{\,\mu\nu}_{\,+}\;+\;\bfpi^{\,\mu\nu}_{\,-}\,\right)\ =\ S^{\,\mu\nu}
$$
$$
{\sqrt{\mbox{\tt D}}}\,\left(\,\bfpi^{\,\mu\nu}_{\,+}\;-\;\bfpi^{\,\mu\nu}_{\,-}\,\right)\ =\
{{\rm i}}\,\varepsilon^{\,\mu\nu\alpha\beta}\,\eta_{\x\alpha}\,k_{\x\beta}
$$
It follows therefrom that we can build up a pair of complex and spacelike
chiral polarization vectors
by means of a constant and spacelike four vector: for example $\epsilon_{\,\nu}=(0,1,1,1)/\sqrt3\,,$
in such a manner that we can set
\ba
\varepsilon^{\,\mu}_{\x\pm}(k)\ \equiv\
\left[\,\frac{{\bf k}^2-(\x\epsilon\cdot k\x)^2}{2\x{\bf k}^2}\,\right]^{-\x1/2}\,
\bfpi^{\,\mu\nu}_{\,\pm}\;\epsilon_{\,\nu}
\ea
which satisfy the orthogonality relations
\ba
-\,\textstyle\frac12\,g_{\,\mu\nu}\;\varepsilon^{\,\mu\,\ast}_{\x\pm}(k)\,\varepsilon^{\,\nu}_{\x\pm}(k)
+ {\rm c.c.}=1\qquad\quad
g_{\,\mu\nu}\,\varepsilon^{\,\mu\,\ast}_{\x\pm}(k)\,\varepsilon^{\,\nu}_{\x\mp}(k)+{\rm c.c.}=0
\ea

as well as the closure relations
\ba
-\,\textstyle\frac12\left[\,
\varepsilon^{\,\mu\,\ast}_{\x+}(k)\,\varepsilon^{\,\nu}_{\x+}(k) +
\varepsilon^{\,\mu\,\ast}_{\x-}(k)\,\varepsilon^{\,\nu}_{\x-}(k)\
+\ {\rm c.c.}\,\right] = {\mbox{\tt D}}^{-\x1}\,{S^{\,\mu\nu}}
\ea
Now we are ready to find the general solution of the free field equations
(\ref{eqmofgauge}) for the special choice $\varkappa=1$ of the St\"uckelberg parameter.
As a matter of fact, from the relationships
(\ref{projectors1}) and (\ref{projectors2}) we readily obtain
\ba
K^{\,\mu}_{\;\;\nu}\,\varepsilon^{\,\nu}_{\x\pm}(k) &=&
\left[\,\delta^{\,\mu}_{\;\;\nu}(\x k^2-m_{\,\gamma}^{\x2}\x)
+ {\sqrt{\mbox{\tt D}}}\,\left(\,\bfpi^{\,\mu}_{\,+\,\nu}\;-\;\bfpi^{\,\mu}_{\,-\,\nu}\,\right)\,\right]
\varepsilon^{\,\mu}_{\x\pm}(k)\no
&=& \left(\x k^2-m_{\,\gamma}^{\x2} \pm\,\sqrt{\mbox{\tt D}} \,\right)\,
\varepsilon^{\,\mu}_{\x\pm}(k)
\ea
which shows that the polarization vectors of positive and negative chirality
respectively are
solutions of the vector field equations if and only if
\ba
k^{\,\mu}_{\,\pm}&=&(\omega_{\,{\bf k}\,,\,\pm}\,,\,{\bf k})\qquad\quad
\omega_{\,{\bf k}\,,\,\pm}=\displaystyle
\sqrt{{\bf k}^2+m_{\,\gamma}^{\x2}\pm\eta\,|\x{\bf k}\x|}
\label{displiv}
\\
\varepsilon^{\,\mu}_{\x\pm}({\bf k},\eta)&=&\varepsilon^{\,\mu}_{\x\pm}(k_{\x\pm})\qquad\quad
\left(\,k^{\x0}_{\x\pm}\;=\;\omega_{\,{\bf k}\,,\,\pm}\,\right)
\ea
Notice that the above pair of LIV chiral polarizations do not coincide at all
with the transverse elliptic polarizations of the conventional Maxwell plane waves.
Nonetheless,
if we introduce the further pair of orthonormal polarization four vectors,
i.e., the temporal and longitudinal polarization real vectors respectively,
\ba
\varepsilon^{\,\mu}_{\,T}\x(k)\;\equiv\;
\frac{k^{\,\mu}}{\sqrt{\,k^2}}\qquad\qquad
(\,k^2>0\,)\\
\varepsilon^{\,\mu}_{\,L}\x(k)\;\equiv\;(\x k^2\x\mbox{\tt D}\,)^{-\,1/2}\left(
k^2\,\eta^{\,\mu} - k^{\,\mu}\,\eta\cdot k\,\right)
\qquad\qquad
(\,k^2>0\,)
\ea
which fulfill by construction
\ba
g_{\,\mu\nu}\;\varepsilon^{\,\mu}_{\,T}\x(k)\,\varepsilon^{\,\nu}_{\,T}\x(k)\ =\ 1\
=\ -\;g_{\,\mu\nu}\;\varepsilon^{\,\mu}_{\,L}\x(k)\,\varepsilon^{\,\nu}_{\,L}\x(k)
\ea
\ba
g_{\,\mu\nu}\,\varepsilon^{\,\mu}_{\,T}\x(k)\,\varepsilon^{\,\nu}_{\,L}\x(k)=
g_{\,\mu\nu}\,\varepsilon^{\,\mu}_{\,T}\x(k)\,\varepsilon^{\,\nu}_{\x\pm}(k)=
g_{\,\mu\nu}\,\varepsilon^{\,\mu}_{\,L}\x(k)\,\varepsilon^{\,\nu}_{\x\pm}(k)=0
\ea
then we have at our disposal $\forall\,k^{\,\mu}$ with $k^2>0$ a complete
orthonormal set of four polarization four vectors : namely,
\ba
\varepsilon^{\,\mu}_{\,A}\x(k)=\left\lbrace
\begin{array}{cc}
{k^{\,\mu}}/{\sqrt{\,k^2}} & {\rm for}\ A=T\\
\left(k^2\,\eta^{\,\mu} - k^{\,\mu}\,\eta\cdot k\,\right)/
\displaystyle\sqrt{k^2\x\mbox{\tt D}} & {\rm for}\ A=L\\
\varepsilon^{\,\mu}_{\x\pm}(k_{\x\pm}) & {\rm for}\ A=\pm
\end{array}
\right.
\qquad\quad
(\,k^2>0\,)
\ea
It follows that if we introduce the $4\times4$ polarization matrix
\ba
g_{\,AB}=g^{\,AB}
\equiv\left\lgroup
\begin{array}{cccc}
1 & 0 & 0 & 0\\
0 & -1 & 0 & 0\\
0 & 0 & -1 & 0\\
0 & 0 & 0 & -1\
\end{array}\right\rgroup
\qquad\quad
(\, A,B=T,L,+,-\,)
\ea
then we can write the full orthogonality and closure relations
\ba
{\textstyle\frac12}\,g_{\,\mu\nu}\;\varepsilon^{\,\mu\,\ast}_{\,A}\x(k)\,\varepsilon^{\,\nu}_{\,B}\x(k)
+ {\rm c.c.}\,=\,g_{\,AB}\qquad\quad
{\textstyle\frac12}\,g^{\,AB}\,\varepsilon^{\,\mu\,\ast}_{\,A}\x(k)\,\varepsilon^{\,\nu}_{\,B}\x(k)
+ {\rm c.c.}\,=\,g^{\,\mu\nu}
\label{orthoclosure}
\ea
It is very important to realize that the on mass shell polarization vector
$\varepsilon^{\,\mu}_{\,-}\x(k_-)$ with $k_-^2=m^{\x2}_{\x\gamma}-\eta\,|\x{\bf k}\x|>0\,,$
is well defined
if and only if the spatial momentum ${\bf k}$ stands below the
momentum cutoff $\Lambda_{\,\gamma}\,,$ i.e. inside the
large momentum sphere
\ba
|\x{\bf k}\x|<\frac{m^{\x2}_{\x\gamma}}{\eta}\equiv\Lambda_{\,\gamma}
\ea
Now, in order to implement the canonical quantization of the LIV massive
vector field for the especially simple choice $\varkappa=1$, it is convenient to introduce
the plane waves according to
\ba
u_{\,{\bf k}\,A}^{\,\nu}\x(x) =
\left[\,(2\pi)^3\,2\omega_{\,{\bf k}\,A}\,\right]^{\x-\x1/2}\,\varepsilon^{\,\nu}_{\,A}\x({\bf k})\
\exp\{-\,{\rm i}\x\omega_{\,{\bf k}\,A}\x x^0+{\rm i}\x{\bf k}\cdot{\bf x}\}
\ea
where
\ba
&&\varepsilon^{\,\nu}_{\,T}\x({\bf k})=\frac{k^{\,\nu}}{m_{\x\gamma}}\qquad\quad
\varepsilon^{\,\nu}_{\,L}\x({\bf k})=
\frac{m_{\x\gamma}^{\x2}\,\eta^{\,\nu} - k^{\,\nu}\x(\eta\cdot k)}{m_{\x\gamma}\,|\x{\bf k}\x|\,\eta}\\
{\rm for}\quad
&&k_0=\omega_{\,{\bf k}\,T}=\omega_{\,{\bf k}\,L}\;=\;
\displaystyle\sqrt{{\bf k}^{\x2}+m^{\x2}_{\x\gamma}}\;\equiv\;\omega_{\,{\bf k}}\\
&&\varepsilon^{\,\nu}_{\,+}\x({\bf k})\equiv\varepsilon^{\,\nu}_{\x+}(k_{\x+})\qquad\quad
\varepsilon^{\,\nu}_{\,-}\x({\bf k})\equiv\varepsilon^{\,\nu}_{\x-}(k_{\x-})\;
\theta\left(\x k_{\x-}^{\x2}\x\right)
\ea
It follows therefrom that the canonical quantization of the free LIV massive vector
field for $\varkappa=1$ takes the form
\ba
A^{\nu}(x)=\sum_{{\bf k}\,,\,A}\,
\left[\,a_{\,{\bf k}\,A}\,u_{\,{\bf k}\,A}^{\,\nu}\x(x) +
a_{\,{\bf k}\,A}^{\,\dagger}\,u_{\,{\bf k}\,A}^{\,\nu\,\ast}\x(x)\,\right]
\ea
where
\ba
\sum_{{\bf k}\,,\,A}\equiv\sum_{A=T,L,\pm}\int{\rm d}\x{\bf k}
\ea
whereas the canonical commutation relations holds true, viz.,
\ba
\left[\,a_{\,{\bf k}\,A}\,,\,a_{\,{\bf p}\,B}^{\,\dagger}\,\right]\;=\;-\,\eta_{\,AB}\,
\delta({\bf k}-{\bf p})
\ea
all the other commutators being equal to zero. According to equation (\ref{eqmofgauge1})
we obtain
\ba
B(x)=\frac{1}{{\rm i}}\int{\rm d}\x{\bf k}\;k_{\x\nu}\left[\,a_{\,{\bf k}\,T}\,\,u_{\,{\bf k}\,T}^{\,\nu}\x(x) -
a_{\,{\bf k}\,T}^{\,\dagger}\,u_{\,{\bf k}\,T}^{\,\nu\,\ast}\x(x)\,\right]_{k_{\x0}\,=\,\omega_{\,{\bf k}}}
\ea
in such a manner that the physical Hilbert space ${\mathfrak H}_{\x\rm phys}$ with positive semi-definite metric,
for the LIV massive $\widetilde\gamma-$photons,
is selected out from the Fock space ${\mathfrak F}$ by means of the subsidiary condition
\ba
B^{\,(-)}(x)\,|\,\rm phys\,\rangle\;=\;0\qquad\quad
\forall\,|\,\rm phys\,\rangle\,\in\,{\mathfrak H}_{\x\rm phys}\subset{\mathfrak F}
\ea
which keeps true even in the presence of the interaction with the spinor field, as described by the
basic Lagrange density (\ref{lagrangian}). On the other side, it turns out that
the LIV massive $\widetilde\gamma-$photons are described by the quantized field
\ba
V^{\x\nu}(x)=
\int{\rm d}\x{\bf k}
\sum_{A\x=\,L,\pm}\left[\,a_{\,{\bf k}\,A}\,u_{\,{\bf k}\,A}^{\,\nu}\x(x) +
a_{\,{\bf k}\,A}^{\,\dagger}\,u_{\,{\bf k}\,A}^{\,\nu\,\ast}\x(x)\,\right]
\ea
with the standard non-vanishing canonical commutation relations
$$
\left[\,a_{\,{\bf k}\,A}\,,\,a_{\,{\bf p}\,B}^{\,\dagger}\,\right]\;=\;\delta_{\,AB}\,
\delta({\bf k}-{\bf p})\qquad\quad A,B=L,\pm
$$
all the other commutators being equal to zero.
Notice that the LIV massive $\widetilde\gamma-$photon 1-particle states
of definite spatial momentum $\bf k$ do exhibit three polarization states :
one linear longitudinal polarization of real vector $\varepsilon^{\,\nu}_{\,L}(k)$
with dispersion relation $k^2=m_{\,\gamma}^{\x2}$
and two chiral transverse states with complex vectors $\varepsilon^{\,\nu}_{\,\pm}(k_{\x\pm})$
and dispersion relations (\ref{eqphot}) and (\ref{displiv}),
the negative chirality states $\varepsilon^{\,\nu}_{\,-}(k_-)$ being well defined
only for $|\x{\bf k}\x|<\Lambda_{\,\gamma}\,\Leftrightarrow\,k_{\x-}^{\x2}>0\,.$

\medskip
If the \AE ther is timelike and isotropic $\eta_\mu = (\eta,0,0,0)\,\propto\,b_\mu = (b,0,0,0)$,
then from eqs.~\gl{eqel}, \gl{eqphot} and \gl{displiv} one can find the following
dispersion relations for the photon chiral polarizations and for fermion normal modes
respectively:
namely,
\ba
&& \omega_{\,{\bf k}\,,\,\pm}^{\,2} = k^2+ m^2_\gamma\pm k\eta\ ,\\
&& \omega_{\,{\bf p}\,,\,\pm}^{\,2} = p^2+ m^2_e+b^2\pm 2bp\ ,\\
&& k_0 = \pm \sqrt{(k\pm{\scriptstyle\frac12}\eta)^2 + m^2_{\;\!\gamma} - {\textstyle\frac14}\;\!\eta^2}
\equiv \pm\,\omega_{\,{\bf k}\,,\,\pm}\ ,
\label{disphot}\\
&& p_0 = \pm \sqrt{(p\pm b)^2 + m^2_e} \equiv \pm\,\omega_{\,{\bf p}\,,\,\pm}\ .
\label{disel}
\ea
in which we have set $k^{\,\mu}=(k_0,{\bf k})\,,\,p^{\,\mu}=(p_0,{\bf p})$
and $k=|{\bf k}|\,,\,p=|{\bf p}|$. The necessary condition for stability is
$ m^2_\gamma \geq \frac14 \eta^2$ but in fact it is not sufficient,
as we shall see further on. Both dispersion relations exhibit a similar pattern of velocities.
Namely, the group velocities for fermions are bounded by the conventional speed of light
$(\,\widehat{\bf a}={\bf a}/|{\bf a}|\,)$
\be
{\bf v}_{\,\pm}\ \equiv\ \nabla_{\bf p}\,\omega_{\,{\bf p}\,,\,\pm}\ =\
\widehat{\bf p}\,(\,p\pm b\,)\,\omega^{\,-\,1}_{\,{\bf p}\,,\,\pm}\ ,
\qquad\quad |\,{\bf v}_{\,\pm}\,|<1
\ee
and the very same for the LIV modified photons with $b \rightarrow \eta/2,\
m^2_e \rightarrow m^2_{\;\!\gamma} - {\textstyle\frac14}\;\!\eta^2$
\be
{\bf u}_{\,\pm}\ \equiv\ \nabla_{\bf k}\,\omega_{\,{\bf k}\,,\,\pm}\ =\
\widehat{\bf k}\,(\,k\pm\eta/2\,)\,\omega^{\,-\,1}_{\,{\bf k}\,,\,\pm}\ ,
\qquad\quad |\,{\bf u}_{\,\pm}\,|<1\ .
\ee
Their unusual feature is the variation in magnitude and sign depending on wave vectors.

On the other side the phase velocities for both particles in the
LIV background are not bounded by the conventional speed of light when
chiralities are chosen in \gl{disphot} and \gl{disel} with negative signs,
\ba
\quad w_{\gamma\,,\,-}\ \equiv\ k^{-1}\,\omega_{\,{\bf k}\,,\,-}\,
\left\{\begin{array}{ll}\geq 1\,,\ & {\rm for}\ k \leq {m^2_\gamma}/{\eta}\ ,\\
<1\,,\ & {\rm for}\ k > {m^2_\gamma}/{\eta}\ ; \end{array}\right.\\
w_{e\,,\,-}\ \equiv\ p^{-1}\,{\omega_{\,{\bf p}\,,\,-}}\,
\left\{\begin{array}{ll}\geq 1\,,\ & {\rm for}\ p \leq (m^2_e + b^2)/2b\ ,\\
<1\,,\ & {\rm for}\ p > (m^2_e + b^2)/2b\ . \end{array}\right.
\ea
In other words, the 1-particle four vectors $k^{\,\mu}$ and $p^{\,\mu}$
for negative chiralities 1-particle states do leave the causality (and further on stability) region when
\be
k^{\,\mu}k_\mu<0\quad{\rm for}\quad|\,{\bf k}\,|> m_{\;\!\gamma}^2/\eta\,,\qquad
p^{\,\mu}p_\mu<0\quad{\rm for}\quad|\,{\bf p}\,|\,>\, m_e^2/2b\,.
\label{physmot}
\ee
In spite of similarity the two bounds are quite different phenomenologically:
whereas the electron mass is estimated to be much larger than any possible background
$b\,,$ there is no evidence for a photon mass at a very stringent limit \cite{Amsler:2008zzb,Kahniashvili:2008va}.
However, if the photon mass is strictly zero, then the phase
propagation of photons with negative chiralities in the presence of CS interaction
is acausal and the latter ones become unstable
for all values of wave vectors -- see below.

Concerning the 1-particle fermion states,
we see that the necessary {causality} requirement
\be
g^{\,\mu\nu}\,p_\mu\,p_\nu = \omega^{\,2}_{\,{\bf p}\,,\,\pm}-\,p^2 \simeq m_e^2\pm 2b\,p\ >\ 0
\ee
is always fulfilled by the upper frequencies $(\omega_{\,{\bf p}\,,\,+})$,
while for the lower frequencies $(\omega_{\,{\bf p}\,,\,-})$ it leads to the
unavoidable physical ultraviolet cutoff for \eg $\widetilde e^{\,\pm}$ particles
\be
p\ <\ \frac{m_{\;\!e}^2}{2b}\ \equiv\ \Lambda_{\;\!e}
\label{fermionuvcutoff}
\ee
that involves negative chirality 1-particle states and positive chirality 1-antiparticle states.
Now we are ready to face the nontrivial problem of the one loop radiative quantum corrections
induced by the electromagnetic minimal coupling between LIV modified fermions and the
Proca-Chern-Simons vector particles.

\section{The 1-loop Photon Self-Energy}
Our aim is to compute the 1-loop induced parity even effective action. For
the sake of simplicity, we refer to the classical spinor
lagrangian density involving only one species of massive fermion with mass
$m\,,$ {{\rm viz}.,}
\begin{equation}
{\cal L}_{{\rm spinor}}=\bar\psi(x)\left({\rm i}\partial\!\!\!/ + e\A(x) - m -\b\,\gamma_5\right)\psi(x)\,,
\label{4.5}
\end{equation}
which leads to the momentum space four dimensional Feynman propagator
\be
S(p,b,m)= \frac{\left[\,p^2+ b^2-m^2+2\left(b\cdot p+m\b\right)\gamma_5\,\right]
\left( \p +m+\b\gamma_5\right)}{4\,\i\left[\,(b\cdot p)^2-m^2b^2\,\right]
-\i\left(p^2+b^2-m^2+i\varepsilon\right)^2}\,.
\label{4.6}
\ee From the
Feynman rules,
the 1-loop photon self-energy \cite{Peskin:1995ev,Itzykson:1980rh,Bogolyubov:1980nc,Faddeev:1980be}
or vacuum polarization tensor, is formally determined to be
\begin{equation}
\Pi^{\,\mu\nu}(k;b,m)\ =\ -\,ie^2\int\frac{\d^{4}p}{(2\pi)^{4}}\
\tr\left\{\gamma^\mu\, S(p)\,\gamma^\nu\, S(p+k)\right\}\,.
\label{4.7}
\end{equation}
However, the above formal expression does exhibit ultraviolet divergencies by
superficial power counting, which have to be properly regularized.
The general structure of the regularized 1-loop photon
self-energy tensor looks as follows
\begin{equation}
\reg \Pi^{\mu\nu}\ =\ \reg \Pi^{\,\mu\nu}_{\,\rm even}\ +\ \reg \Pi^{\,\mu\nu}_{\,\rm odd}\ .
\label{4.8}
\end{equation}
The 1-loop parity odd part has been evaluated \cite{Andrianov:2001zj}
for a vanishing external momentum and reads
\ba
\reg \Pi^{\ \mu\nu}_{\,\rm odd}=
\frac{\i\,e^2}{2\pi^2}\,\epsilon^{\,\mu\nu\rho\sigma} b_\rho k_\sigma
\qquad\quad(b^2\ll m^2)\ ,
\ea
which corresponds to the induced Chern-Simons coefficient
\be
\Delta\,\eta_{\mu}\;=\;-\,\frac{2\alpha}{\pi}\,b_{\mu}\qquad\qquad(\alpha=e^2/4\pi)\ .
\label{chsim}
\ee
The regularization independence as well as the physical meaning of the
parity odd part of the polarization tensor
have been  discussed in refs. ~\cite{Andrianov:2001zj,Ebert:2004pq,Brito:2008ec}.
It has been proven that the dimensional regularization
properly applied to
the parity odd part of polarization tensor \eqref{4.7} with propagators
\eqref{4.6} was able
to reproduce the same induced Chern-Simons coefficient \eqref{chsim} as it was
obtained with the
physically motivated cutoff \eqref{physmot} from Section 2. This result is
accounted for
by the fact that the above CS coefficient is finite and not screened by a
power like or logarithmic
divergence. For the CPT and parity even part induced by LIV the relationship
between two regularizations
is more subtle as the quadratic divergence in the Lorentz invariant part is
dominant and screens
the subdominant effects of LIV contributions. These subtleties will be considered  below.

\subsection{Induced Photon Mass (IPM) in Dimensional Regularization} 
The regularized expression of the 1-loop photon self-energy tensor
is thereby given
\begin{equation}
i\,\reg \Pi^{\,\mu\nu}(k\,;b,m,\mu)\ \equiv\ e^2 \mu^{4-2\omega}\int\frac{\d^{2\omega}p}
{(2\pi)^{2\omega}}\ \tr \left\{\gamma^\mu\, S(p)\,\gamma^\nu\,
S(p+k)\right\},
\label{5.1}
\end{equation}
where dimensional regularization is employed to give a meaning to the
loop integral, which appears by power counting to be superficially
quadratically divergent in four dimensions.  Notice that the
Dirac matrices involved in the regularized loop
integral~(\ref{5.1}) have now to be understood and treated according to
the algebraically consistent general algorithm \cite{'tHooft:1972fi,Breitenlohner:1977hr,Breitenlohner:1975hg}.

Dimensional regularization is supposed to fulfill gauge invariance.
This is certainly true for a Lorentz covariant field theory model.
When Lorentz covariance is broken, there are at least two
different ways of implementing dimensional regularization \cite{Andrianov:2001zj}.
Both of them do indeed respect the gauge invariance for the parity odd part of the
effective action, up to the 1-loop approximation, although
leading to two different gauge invariant results.
On the contrary, to keep
the gauge invariance is much more subtle for the parity even part
and, eventually, it does no longer hold true.
The details of the calculation can be found in Appendix A.

Here below we proceed to
the 1-loop radiatively induced generation of a photon mass and the corresponding breakdown of the gauge invariance
owing to the presence of a constant temporal axial vector \AE ther background
coupled with the fermionic matter.
After lengthy calculations presented in Appendix B we find the following result for the induced photon mass in a LIV background,
\be
\reg\ \Delta \Pi_{\,{\rm even}}^{\,\mu\nu}\ =\
{2\alpha\over 3\pi}\ b^2\,\bar g^{\,\mu\nu}\ .
\label{finitephotonmass}
\ee
The important meaning of the above result is that, in the presence of an explicit breaking of the Lorentz invariance in the fermionic sector, the gauge symmetry of the abelian vector field may be lost, owing to anomalous radiative quantum corrections,
so that an induced tiny Proca mass term for the $\widetilde\gamma-$photon field does arise in the effective lagrangian,
that is
\be
\frac12\,\Delta m_\gamma^2\,A_\mu\,A^\mu\ , \qquad\quad \Delta m_\gamma^2 = {2\alpha\over 3\pi}\ b^2\ .
\ee
This radiatively generated Proca mass term
\footnote{Notice that the radiatively generated photon mass term was already reported in our previous
paper \cite{Alfaro:2006dd}  but with erroneous sign, \ie a tachyonic imaginary mass instead of a real Proca mass.}
for $\widetilde\gamma-$photons accompanies
the introduction of a general bare photon mass which is necessary
to make the whole formulation of the LIV QED fully consistent.

\subsection{IPM with the Physical UV Cutoff} 

It is instructive to repeat the calculation of the radiatively generated photon mass in the physical cutoff regularization.
To this concern we recall that the large ultraviolet cutoff $\Lambda$,
for 1-particle states of negative helicity as well as 1-antiparticle states of positive helicity with momentum
${\bf p}$ and mass $m$, is provided by eq.~(\ref{fermionuvcutoff})
that is $|\,{\bf p}\,|<\Lambda\simeq m^2/2b$, with $b=b_0$.
In accordance with our previous notations
\be
\reg \Pi_{\,{\rm cov}}^{\,\mu\nu}(k,m)=\left(\,k^2\,g^{\,\mu\nu}-k^{\,\mu}\,k^{\,\nu}\,\right)\,
\reg \Pi_{\,{\rm cov}}(k^2,m^2)+\reg\widetilde\Pi_{\,{\rm cov}}^{\,\mu\nu}(\Lambda^2,m^2)
\ee
we have
\ba
\reg\widetilde\Pi_{\,{\rm cov}}^{\,\mu\nu}(m,\Lambda)&=&
{\i\,e^2\over4\pi^4}\int {\rm d}^4\;\!p\ \theta(\Lambda^2-{\bf p}^2)\, {N_0^{\mu\nu}(p,m)}\,{D^2(p,m)}\no &=&{\i\,\alpha\over\pi^3}\int{\rm d}^4\;\!p\ \theta(\Lambda^2-{\bf p}^2)\, \frac{2p^\mu\,p^\nu-(p^2-m^2)\,g^{\,\mu\nu}}{(p^2-m^2+i\varepsilon)^2}\no
&\equiv& -\,g^{\,\mu\nu}\,m^2\,\reg\Pi_{\,{\rm cov}}(\Lambda/m)\,\ ,
\ea
\ba
\reg\ \Delta \Pi_{\,{\rm even}}^{\,\mu\nu}(b,m,\Lambda) =
2I_0^{\mu\nu}(b,m)+J_0^{\,\mu\nu}(b,m) \equiv A\,g^{\,\mu\nu}\,b^2 + B\,b^{\,\mu}b^{\,\nu}
\ea
and suppose the constant vector breaking Lorentz symmetry to be purely temporal,
that is $b^{\,\mu}=(b,0,0,0)\,.$ Then, on the one side we find
\ba
\reg\Pi_{\,{\rm cov}} &=& \frac{\i\,\alpha}{2m^2\pi^3}\int{\rm d}^4\;\!p\
\theta(\Lambda^2-{\bf p}^2)\, \frac{p^2-2m^2}{(p^2-m^2+i\varepsilon)^2}
\label{picov}
\ea
\ba
\reg\ \Delta \Pi_{\,{\rm even}}^{\,\imath\jmath}(b,m,\Lambda)\;\delta_{\,\imath\jmath}\,
&=& -\,3A\,b^2
\ea
On the other side we obtain
\ba
\reg\Pi_{\,{\rm cov}} &=& {2\,\i\,\alpha\over m^2\pi^2}\int_0^{\,\Lambda}
{\rm d}p\,p^2\int_{-\infty}^\infty {\rm d}p_0\ \left[\,\frac{1}{D}-\frac{m^2}{D^2}\,\right]=
{\alpha\over\pi\;\!m^2}\int_0^{\,\Lambda} {\rm d}p\
\left({2p^2\over\omega_{\,p}}+{p^2m^2\over\omega^{\,3}_{\,p}}\right)
\ea
with
\be
D\equiv p_0^2-{\bf p}^2-m^2+\i\ve=(p_0-\omega_{\,p}+\i\varepsilon)\,
(p_0+\omega_{\,p}-\i\varepsilon)\ ,\qquad\quad \omega_{\,p}=\sqrt{p^2+m^2}\ .
\ee
After setting $\epsilon\,\equiv\,m/\Lambda\simeq2b/m$ we get
\ba
\reg\Pi_{\,{\rm cov}}&=& {\alpha\over\pi\epsilon^2}\, (\,1+\epsilon^2\,)^{-1/2}\no
&=& {\alpha\over\pi\epsilon^2}\left(1-\frac12\,\epsilon^2+\frac38\,\epsilon^4\right)+O(\epsilon^4) \qquad\qquad\left[\,\epsilon={m\over\Lambda}\simeq{2b\over m}\ll 1\,\right]
\label{covpol_0}
\ea
which means that the divergent photon mass term arising from the covariant piece
of the vacuum polarization tensor within the ultraviolet physical cutoff regularization
becomes
\ba
\reg\widetilde\Pi^{\,\mu\nu}_{\,{\rm cov}}(m,\Lambda)=
g^{\,\mu\nu}\,m^2\;{\alpha\over\pi}\left(-\,{1\over\epsilon^2} +
\frac12 - \frac38\,\epsilon^2\right)+O(\epsilon^4)
\ea

\bigskip
Next, the piece of 1-loop vacuum polarization tensor  which in principle
violate Lorentz invariance can be obtained as follows. First we have
\ba
&& -\,2\delta_{\,\imath\jmath}\,I_0^{\,\imath\jmath}(b,m)\ =\no
&& {16\,\i\,e^2b^2\over(2\pi)^4}\int{\rm d}^4\;\!p\
\theta(\Lambda^2-{\bf p}^2)\, \left[\,2{\bf p}^2 + 3(p_0^2 - {\bf p}^2 - m^2)\,\right]\;
\frac{m^2 - p_0^2 - {\bf p}^2}
{\left(\,p_0^2 -{\bf p}^2 - m^2 + \i\varepsilon\,\right)^4}\no
 &=&
{e^2b^2\over{\rm i}\,\pi^3}\int_0^{\,\Lambda} {\rm d}p\int_{-\infty}^\infty {\rm d}p_0\
\left({16p^6\over D^4}+{32p^4\over D^3}+{12p^2\over D^2}\right)\ .
\ea
Moreover we get
\ba
\delta_{\,\imath\jmath}\,J_0^{\,\imath\jmath}(b,m) &=&
{e^2b^2\over\i\,\pi^3}\int_0^{\,\Lambda} {\rm d}p\int_{-\infty}^\infty {\rm d}p_0\
\left({3p^2\over D^2}-{8p^6\over D^{\,4}}\right) ,
\ea
so that we can finally write
\ba
A = -\,{\i\,e^2\over3\pi^3}\int_0^{\,\Lambda} {\rm dp}\int_{-\infty}^\infty {\rm d}p_0\
\left(24\,{p^6\over D^4}+32\,{p^4\over D^3}+9\,{p^2\over D^2}\right)\ .
\ea
and by performing first the energy integration
\ba
A = {e^2\over2\pi^2}\int_0^{\,\Lambda} \d p\ \left({5}\,{p^6\over\omega_{\,p}^{\,7}}-8\,
{p^4\over \omega_{\,p}^{\,5}}+{3}\,{p^2\over\omega_{\,p}^{\,3}}\right)
\ea
and then momentum integration \cite{30} we finally obtain
\ba
A &=&
{2\alpha\over\pi}\,\cdot\,{m^2\over\Lambda^2}\left(1+{m^2\over\Lambda^2}\right)^{-\,\frac52}\ .
\ea
Next we turn to the calculation of the coefficient $B\,.$ To this concern we find
\be
g_{\,\mu\nu}\,\reg\ \Delta \Pi_{\,{\rm even}}^{\,\mu\nu}(b,m,\Lambda) = (4A+B)\,b^2
\ee
and contraction with the metric tensor gives
\ba
g_{\,\mu\nu}\,[\,2I_0^{\mu\nu}(b,m)+J_0^{\,\mu\nu}(b,m)\,]
={3e^2b^2\over2\pi^2}\,m^2\int_0^\Lambda \d p\
\left(3\,{p^2\over\omega_{\,p}^{\,5}} - 5\,{p^4\over\omega_{\,p}^{\,7}}\right)
\ea
It follows that we have
\ba
4A+B\ =\ {3e^2m^2\over2\pi^2}\int_0^\Lambda \d p\
\left(3\,{p^2\over\omega_{\,p}^{\,5}} - 5\,{p^4\over\omega_{\,p}^{\,7}}\right)
\ea
and from \cite{30} we eventually obtain
\ba
4A+B\ =\ {3e^2\over2\pi^2}\left({\Lambda^3\over u^3} - {\Lambda^5\over u^5}\right)
\ea
and consequently
\ba
B\ =\ -\,{2\alpha\over\pi}\,\cdot\,{m^2\over\Lambda^2}\left(1+{m^2\over\Lambda^2}\right)^{-\,\frac52}\
=\ -\,A\ .
\ea

Hence, the physical high momentum fermion cutoff leads to the following result
for the vacuum polarization tensor in the limit of a null photon momentum :
namely,
\ba
\lim_{k\,\to\,0}\ \reg\ \Delta \Pi_{\,{\rm even}}^{\,\mu\nu}(k,b,m,\Lambda) ={2\alpha\over\pi}\,\cdot\,{m^2\over\Lambda^2}\left(1+{m^2\over\Lambda^2}\right)^{-\,\frac52}\, \left(b^2\,g^{\,\mu\nu}-b^{\,\mu}\,b^{\,\nu}\right)
\ea
which corresponds to a {\em bona fide} vanishing contribution to the photon mass
in the presence of a very large momentum  cutoff for spinor matter.

To sum up, we see that in order to remove the previously obtained divergent photon mass term for
$b^2>0$ one has to introduce the Proca mass Lorentz invariant counterterm
\ba
{\mathcal L}_{\,\rm c.t.} = \frac12\,Z_\gamma\,m^2\,A_\mu\,A^\mu
\ea
in which
\ba
Z_\gamma={\alpha\over\pi}\left({1\over \epsilon^2} - F_\gamma\right)+O(\epsilon^4)
\ea
where $F_\gamma(\epsilon)$ is an arbitrary function analytic for $\epsilon\to 0\,.$
A comparison with the result (\ref{finitephotonmass}) fix the finite part
of the counterterm to be given by
\be
F_\gamma(\epsilon)=\frac12 - {13\over24}\,\epsilon^2\ +\ O(\epsilon^4) .
\ee

\subsection{IPM at Finite Temperature}

According to refs.~\cite{Zhukovsky:2005iu,Ebert:2004pq,Brito:2008ec}, it is  a consistency check to get the radiatively induced photon mass term at a finite temperature $T\,.$
To this concern, let us perform the Wick rotation with the conventional substitutions
\ba
{1\over(2\pi)^4}\int \d^{\,4\,}p\ \theta(\Lambda^2-{\bf p}^2)
&\rightarrow& \frac{i}{\beta}\ \sum_{n\,=\,-\,\infty}^\infty\int{\d{\bf p}\over(2\pi)^3}\
\theta(\Lambda^2-{\bf p}^2)\no
p_0 &\rightarrow& i\,\omega_n\ =\ \frac{\pi i}{\beta}\,(2n+1)\qquad\quad
n\in {\mathbb Z}
\ea
where $\beta=1/kT\,,\ k$ being the Boltzmann constant. Hence we obtain
\ba
m^2\,\Pi_{\,\beta} &=& {4\alpha\over \pi\beta}\int_0^{\,\Lambda}
\d p\,p^2 \sum_{n\,=\,-\,\infty}^\infty\ \left(\frac{1}{D_1}+\frac{m^2}{D_1^2}\right)\ ,
\ea
\ba A_\beta = {8\alpha\over3\pi\beta}\sum_{n\,=\,-\,\infty}^\infty\
\int_0^{\,\Lambda} \d p\,p^2\ \left(24\,{p^4\over D_1^4}-32\,{p^2\over D_1^3}
+ {9\over D_1^2}\right)\ , \ea with \ba D_1\ =\ \omega_n^2+m^2+p^2\ =\
{\pi^2\over\beta^2}\,(2n+1)^2+p^2+m^2\ .
\ea
Now, if we suitably set
\be
D_z=\omega_n^2+m^2+zp^2\equiv\,a+zp^2\qquad\quad[\,az>0\,]
\ee
we can rewrite the above expression in the form
\ba
m^2\,\Pi_{\,\beta} = {4\alpha\over \pi\beta} \sum_{n\,=\,-\,\infty}^\infty\
\left[\,{\partial\over\partial a}\int_0^{\,\Lambda} \d p\,p^2\,\ln D_1
- \lim_{z\,\to\,1}\ {\d\over \d z}\int_0^{\,\Lambda}\d p\ {m^2\over D_z}\,\right]\ ,\\
A_\beta = -\,{8\alpha\over3\pi\beta}\sum_{n\,=\,-\,\infty}^\infty\ \lim_{z\,\to\,1}\
\left(9\,{\d\over \d z}+16\,{\d^{\,2}\over \d z^2}+4\,{\d^{\,3}\over \d z^3}\right)
\int_0^{\,\Lambda}{\d p\over D_z}\ .
\ea

From \cite{30} we get
\ba
A_\beta &=& -\,{8\alpha\over3\pi\beta}\sum_{n\,=\,-\,\infty}^\infty\
\lim_{z\,\to\,1}\ \left(9\,{\d\over \d z}+16\,{\d^{\,2}\over \d z^{2}}+4\,
{\d^{\,3}\over\d z^{3}}\right)\,f(z)\ ,\no
f(z) &=& \int_0^{\,\Lambda}{\d p\over D_z}\ =\
{1\over\surd(az)}\,{\rm arctg}\left(\Lambda\,\sqrt{{z\over a}}\,\right)\ ,
\ea
\ba
{\partial\over\partial a}\int_0^{\,\Lambda} \d p\,p^2\,\ln D_1 =
\Lambda - {\sqrt{\,a}}\,{\rm arctg}{\Lambda\over\surd\, a}\ ,
\ea
in such a manner that we eventually find for $a=\omega_n^2+m^2$
\ba
f(1) &=& {1\over\surd a}\,{\rm arctg}{\Lambda\over\surd a} =
{1\over\surd(\omega_n^2+m^2)}\,{\rm arctg}\,\frac{\Lambda}{\surd(\omega_n^2+m^2)}\no
{\d f\over \d z}&=&-\,{1\over2z}\,\left[\,f(z)-{\Lambda\over
    a+z\Lambda^2}\,\right]\ ,\no
{\d^{\,2}f\over \d z^2}&=& {3\over4z^2}\,\left[\,f(z)-{\Lambda\over
    a+z\Lambda^2}\,\right]-{1\over2z}\,\cdot\,{\Lambda^3\over(
  a+z\Lambda^2)^2}\ ,\no
{\d^{\,3}f\over \d z^3}&=& -\,{15\over8z^3}\,\left[\,f(z)-{\Lambda\over
    a+z\Lambda^2}\,\right]+{5a\Lambda^3+(4+5z)\Lambda^5\over4z^2(
  a+z\Lambda^2)^3}\ ,
\ea
As a preliminary check of the above formul\ae, let us first reproduce the value of
the integral which occurs in the calculation of the temperature dependent parity
odd part for the photon polarization tensor. To this aim,
consider the integral \cite{Ebert:2004pq}
\ba
I_{\rm odd}&=&\int_0^\Lambda \d p\,p^2\ \frac{3\omega_n^2+3m^2-p^2}{(\omega_n^2+m^2+p^2)^3}\no
&=&\lim_{z\,\to\,1}\ \left(-\,3\,{\d\over \d z}-2\,{\d^{\,2}\over
    \d z^2}\right)\int_0^\Lambda {\d p\over D_z}\no
&=& {3\over2}\,\left[\,f(1)-{\Lambda\over
    a+\Lambda^2}\,\right]-{3\over2}\,\left[\,f(1)-{\Lambda\over
    a+\Lambda^2}\,\right]+{\Lambda^3\over(a+\Lambda^2)^2}\no
&=&{\Lambda^3\over(\omega_n^2+m^2+\Lambda^2)^2}\ =\
{\Lambda^3\,(\beta/\pi)^4\over[\,(2n+1)^2+(\beta/\pi)^2(\Lambda^2+m^2)\,]^2}
\ea
which is in perfect agreement with eq.~(11) of ref.~\cite{Ebert:2004pq}~.
Moreover, after setting
\be
\eta\ \equiv\ {\beta\over\pi}\,\displaystyle\sqrt{m^2+\Lambda^2}\,,
\ee
by taking the sum from \cite{30} we have
\ba
&& \sum_{n\,=\,-\,\infty}^\infty\ {2e^2\,\Lambda^3\,\beta^3/
\pi^6\over[\,(2n+1)^2+(\beta/\pi)^2(\Lambda^2+m^2)\,]^2}\no
&=&{e^2\,\Lambda^3\beta^3\over4\eta^3\pi^5}\,\left(2\tanh{\pi\eta\over2}
-\,{\pi\eta}\,{\rm sech}^2{\pi\eta\over2}\right)\no
&\approx& {\alpha\over\pi}\left[\,2\tanh{\pi\eta\over2}-\pi\eta+\pi\eta\,\tanh^2{\pi\eta\over2}\,\right]=
\left\lbrace
\begin{array}{cc} 0\ , & \qquad {\rm for}\  \beta=0\\
2\alpha/\pi\ , & \qquad {\rm for}\  T=0\end{array} \right.
\ea
which is again in accordance with eq.~(11) of ref.~\cite{Zhukovsky:2005iu}.

\bigskip
Turning back to the parity even part of vacuum polarization we find for $a=\omega_n^2+m^2$
\ba
m^2\,\Pi_{\,\beta} &=& {4\alpha\over \pi\beta}\sum_{n\,=\,-\,\infty}^\infty\
\left\lbrace\Lambda - \sqrt{\,a}\,{\rm arctg}{\Lambda\over\surd\, a}\right.\no
&+& \left.\frac12\,m^2\left[\,{1\over\surd a}\,{\rm arctg}{\Lambda\over\surd a}
- {\Lambda\over a+\Lambda^2}\,\right]\right\rbrace
\ea
\ba
A_\beta &=& {8\alpha\Lambda^3\over3\pi\beta}\sum_{n\,=\,-\,\infty}^\infty\
{3a - \Lambda^2\over(a+\Lambda^2)^3}\no
&=& {8\alpha\Lambda^3\over3\pi\beta}\sum_{n\,=\,-\,\infty}^\infty\ \left\lbrace
{3\over(a+\Lambda^2)^2} - {4\Lambda^2\over(a+\Lambda^2)^3}\right\rbrace\no
&=&{8\alpha\over3\pi^2}\left\lbrace 3\left({\beta\Lambda\over\pi}\right)^3\,\Sigma_2\
-\ 4\left({\beta\Lambda\over\pi}\right)^5\,\Sigma_3\right\rbrace\ ,
\ea
in such a manner that we come to the limiting values
\ba
\lim_{\,\beta\,\to\,0}\,A_\beta =0\ ,\qquad\quad
\lim_{\,\beta\,\to\,\infty}\,A_\beta = {2\alpha\over\pi}\,\cdot\,
{m^2\over\Lambda^2}\left(1+{m^2\over\Lambda^2}\right)^{-\,\frac52}\ ,
\ea
in perfect agreement with our previous calculation.
On the other side we have that the following integral representations hold true: namely,
\ba
\Lambda - \sqrt{\,a}\,{\rm arctg}{\Lambda\over\surd\, a} =
\int_0^\Lambda \d y\ {y^2\over y^2+a}\ ,\\
{1\over\surd a}\,{\rm arctg}{\Lambda\over\surd a} = \int_0^\Lambda {\d y\over y^2+a}\ .
\ea

Then we find, after setting $Y(y)=(\,\beta/\pi)\displaystyle\sqrt{m^2+y^2}\,,\
\eta=(\,\beta/\pi)\displaystyle\sqrt{m^2+\Lambda^2}\,,$
\ba
m^2\,\Pi_{\,\beta} &=& {4\alpha\over \pi\beta}\int_0^\Lambda
\d y\sum_{n\,=\,-\,\infty}^\infty\ \left\lbrace{y^2+m^2/2\over y^2+\omega_n^2+m^2} -
{m^2/2\over\omega_n^2+m^2+\Lambda^2}\right\rbrace\no
&=& {4\alpha\beta\over \pi^3}\int_0^\Lambda \d y\sum_{n\,=\,0}^\infty\ \left\lbrace{2y^2+m^2\over(2n+1)^2+Y^2}-{m^2\over(2n+1)^2+\eta^2}\right\rbrace\no
&=& {\alpha\beta\over\pi^2}\left\lbrace\int_0^\Lambda {\d y\over Y}\
(2y^2+m^2)\tanh{\pi Y\over2}-{m^2\Lambda\over\eta}\,\tanh{\pi\eta\over2}\right\rbrace
\ea
Hence we finally get
\ba
\Pi_{\,\beta} &=& \left\lbrace
\begin{array}{cc} 0\ , & \qquad {\rm for}\  \beta=0\\
\Pi_{\,\infty} \ , & \qquad {\rm for}\  T=0\end{array} \right.
\ea
where
\ba
\Pi_{\,\infty}\ =\ \left({\alpha\over\pi}\right)\;{1\over\epsilon^2}\,(\,1+\epsilon^2\,)^{-\,1/2}\ ,
\qquad\quad\epsilon={m\over\Lambda}\ ,
\ea
in full agreement once again with our previous equation (\ref{covpol_0}).

\subsection{Discussion of the Physical Meaning }
Let us discuss the physical meaning of the above result. The large momentum physical bound
(\ref{fermionuvcutoff}) for spinor matter is necessarily provided by the fermion operator
$\bar\psi\gamma^\mu\gamma_5\psi$, which is CPT odd and of
mass dimension three, coupled to the \AE ther's temporal vector
$b_\mu=(b,0,0,0)$.

This means that the LIV 1-particle states of momentum ${\bf p}$,
for charged massive spinor fields, can be understood as a complete set of
stable asymptotic states for each polarization, if and only if the momenta stay
below the large energy momentum physical bound $\Lambda\,.$
For a universal flavor independent \AE ther, a conservative ultraviolet momentum cutoff
for $\widetilde e^{\,\mp}$ particles is expressed by
\ba
|\,{\bf p}\,|<\Lambda_{\;\!e}\simeq {m^2_e\over2b}<10^{26}\ {\rm eV}\ .
\label{stabilitybound}
\ea
Once the existence of such a large physical ultraviolet cutoff has been acknowledged for spinor matter,
then the radiatively induced parameters can be determined. In particular, the 1-loop radiatively induced
Chern-Simons coefficient and $\widetilde\gamma-$photon mass are given
by \cite{Andrianov:2001zj,Ebert:2004pq,Brito:2008ec}
\ba
\Delta\,\eta^{\,\nu}\,=\,-\;{2\alpha\over\pi}\,b^{\,\nu}\ ,\qquad\quad
\Delta\,m_{\;\!\gamma}^2={2\alpha\over3\pi}\,b^{\,\mu}b_{\,\mu}\ ,
\ea
where $\alpha$ is the fine structure constant,
as a result of $\overline{DR}$ dimensional regularization.
We remark that the above induced Chern-Simons coefficient turns out to be finite
and it is not screened
by any power-like or logarithmic divergence.
It is thereby uniquely reproduced in the physical cutoff regularization
as well as with in the finite temperature method.
For the parity even part induced by LIV the relationship between
the physically motivated regularizations is less definite, as the quadratic divergence
in the Lorentz invariant part is dominant and screens the subleading LIV effects.
In this case the proper subtraction of the Lorentz invariant divergence
is provided by the requirement of universality, i.e. coincidence with
$\overline{DR}$ dimensional regularization.
\subsection{{LIV Vacuum Polarization at the Leading Order}}
%
After a straightforward but tedious calculation, it is possible to prove
that the $k-$dependent part of the Lorentz
invariance violating $O(b^2)$ correction to the 1-loop vacuum polarization tensor
does actually fulfill transversality with respect to the external momentum
$k_\mu\,,$ as well as to the \AE ther's vector $b_\mu\,.$
It is convenient to write \cite{Alfaro:2006dd} the
1-loop $\widetilde\gamma-$photon self--energy, or LIV QED vacuum polarization tensor,
up to the quadratic approximation in the Lorentz symmetry breaking four vector
$b_\mu\,.$  It consists in the sum of the Lorentz covariant part and of
the radiatively generated Lorentz symmetry breaking part: namely,
\ba
\reg\,\Pi^{\,\mu\nu}(b,k,m)&\approx&\reg\,\Pi_{\,{\rm cov}}^{\,\mu\nu}(k,m)
+ \Delta \Pi_{\,{\rm even}}^{\,\mu\nu}(b,k,m)
+ \Delta \Pi_{\,{\rm odd}}^{\,\mu\nu}(b,k,m)\no
\reg\,\Pi_{\,{\rm cov}}^{\,\mu\nu}(k,m)&=&
(\,k^2\,g^{\,\mu\nu} - k^\mu k^\nu\,)\,\reg\,\Pi(k^2)\\
\Delta\Pi_{\,{\rm even}}^{\,\mu\nu}(b,k,m) &\approx&
{\alpha\over \pi}\,\left\{\frac23\,b^2\,g^{\,\mu\nu}
- A(b,k,m)\,S^{\,\mu\nu}\right\}\no
\Delta\Pi_{\,{\rm odd}}^{\,\mu\nu}(b,k,m) &\approx&
2{\rm i}\,\frac{\alpha}{\pi}\;\varepsilon^{\,\mu\nu\rho\sigma}\,b_\rho\,k_\sigma
\ea
where
\be
S^{\,\mu\nu} = [\,(b\cdot k)^2 - b^2\,k^2\,]\,\bar g^{\,\mu\nu}
- (b\cdot k)\,(b^\mu\,k^\nu + b^\nu\,k^\mu)
+ b^2\,k^\mu k^\nu + k^2\,b^\mu b^\nu
\ee
whereas, using dimensional regularization with $2\omega$ spacetime dimensions,
\be
\reg\,\Pi(k^2) = -\,\frac{\alpha}{3\pi}\left(\frac{1}{2-\omega}-{\bf C}-\ln\,\frac{m^2}{4\pi\mu^2}
+\frac{k^2}{5m^2}\right)+O\left(\frac{k^2}{m^2}\right)^2
\ee
The form factor $A(b,k,m)$ can be calculated by looking \eg
at the coefficient of the tensor
\[
(b\cdot k)\,(b^\mu\,k^\nu + b^\nu\,k^\mu)
\]
which can be readily extracted from the integral
of equation (\ref{I3}) in the Appendix A.
We find
\ba
&& {\alpha\over \pi}\,b\cdot k\,(b^\mu k^\nu+b^\nu k^\mu)\,A(b,k,m)\ =\no
&-& 8ie^2\int_p \frac{b\cdot(p+k)(p^2-m^2)+b\cdot p\,[\,(p+k)^2-m^2\,]}
{(p^2-m^2+i\varepsilon)^2[\,(p+k)^2-m^2+i\varepsilon\,]^2}\no
&\times& [\,b^\mu p^\nu+b^\nu p^\mu + b^\mu (p+k)^\nu+b^\nu (p+k)^\mu\,]\no
&-& 16ie^2\,m^2\,b\cdot k\,(b^\mu k^\nu+b^\nu k^\mu)\,I(2,2)\no
&& =\ -\ {4\alpha\,m^2\over \pi(k^2)^2}\,b\cdot k\,(b^\mu k^\nu+b^\nu k^\mu)
   \int_0^1 {dx\over R^2}\ x\,(1-x)\ .
\ea
where $R\equiv\,x^2-x+m^2/k^2\,.$ From ref.~\cite{30} we obtain
\ba A(b,k,m) &=&  -\ {4m^2\over (k^2)^2}\left\lbrace \frac{2k^2}{4m^2-k^2}
+ \frac{2m^2-k^2}{4m^2-k^2} \int_0^1 {dx\over R}\right\rbrace\no
&=& \frac{-\,8m^2}{k^2\,(4m^2-k^2)}\no &-& \frac{16m^2\,(2m^2-k^2)}
{[\,-\,k^2\,(4m^2-k^2)\,]^{\,3/2}}\,\cdot\, {\rm Arcth}\,\sqrt{\frac{-k^2}{4m^2-k^2}}\ .
\ea
By performing the Wick rotation in turning to the Euclidean formulation we get
\ba
A(b,k_E,m) &=& \frac{8m^2}{k_E^2\,(4m^2+k_E^2)}\no
&-& \frac{16m^2\,(2m^2+k_E^2)}{[\,k_E^2\,(4m^2+k_E^2)\,]^{\,3/2}}\,\cdot\,
{\rm Arcth}\,\sqrt{\frac{k_E^2}{4m^2+k_E^2}}\no
&=& \left\lbrace (4m^2+2k_E^2)\,{\rm Arcth}\,\sqrt{\frac{k_E^2}{4m^2+k_E^2}} -
\sqrt{k_E^2(4m^2+k_E^2)}\right\rbrace\no
&\times& \frac{8m^2}{[\,k_E^2\,(4m^2+k_E^2)\,]^{\,3/2}}\ .
\ea
We notice that in the limit of a null external momentum we find
\ba
\lim_{\,k\to 0}\;k^{\,2}\,A(b,k,m)\,=\;\frac{-\,2}{3m^2}
\ea
which entails that the order of magnitude of the Lorentz invariance violating
corrections to the photon polarization tensor turns out to be
vanishingly small, {\rm viz.}
$b^2/m^2\simeq 4.5\times 10^{-42}\,.$
It follows therefrom\footnote{According to our conventions,
the 2-point proper vertex turns out to be
$\Gamma^{\,\mu\nu}=-\,k^2\,g^{\,\mu\nu} + k^\mu k^\nu+\Pi^{\,\mu\nu}\,.$}
that the momentum space effective kinetic operator
takes the form
\ba
{\textstyle\frac12}\,(\,-\,k^2\,g^{\,\mu\nu} + k^\mu k^\nu\,)\,[\,1-\Pi(0)\,] +
\frac{\alpha}{3\pi}\left(\,b^2\,g^{\,\mu\nu}
+ m^{-2}\;S^{\,\mu\nu}\,\right)
+ \frac{\alpha}{\pi}\;i\,\varepsilon^{\,\mu\nu\rho\sigma}\,b_\rho\,k_\sigma
\ea
which drives to the action and lagrangian counterterm in the BPHZ subtraction scheme
\ba
{\mathcal A}_{\,\rm c.t.} &=& {\textstyle\frac12}\,\Pi(0)\int\frac{{\rm d}^4k}{(2\pi)^4}\
\widetilde A_\mu(k)(\,-\,k^2\,g^{\,\mu\nu} + k^\mu k^\nu\,)\,
\widetilde A_\nu(-k) \\
{\mathcal L}_{\,\rm c.t.} &=& -\,{\textstyle\frac14}\,(Z_3-1)\,F^{\,\mu\nu}(x)\,F_{\,\mu\nu}(x)\\
Z_3 &=& 1+\Pi(0)=
1 + \frac{\alpha}{3\pi}\left(-\,\frac{1}{\epsilon}+{\bf C}+\ln\,\frac{m^2}{4\pi\mu^2}\right)
\ea
Moreover, we retrieve \cite{Alfaro:2006dd} the finite lowest order radiatively induced local effective lagrangian
for the LIV $\widetilde\gamma-$photon : namely,
\ba
{\cal L}_{\,\rm eff} = &-& \frac14\,F^{\,\mu\nu} F_{\mu\nu} \Big(1+{2\alpha b^2\over3\pi m^2}\Big)
+ {\alpha b^2\over 3\pi}\,A^\mu A_\mu\no
&+& {\alpha\over 3\pi m^2}\ b_\nu b^{\,\rho} F^{\,\nu\lambda} F_{\rho\lambda}  - {\alpha\over2\pi}\,b_\lambda\,A_\mu\,\epsilon^{\,\lambda\mu\rho\sigma} F_{\rho\sigma}\no
&=& -\,{\textstyle\frac14}\,(1+\ve)\,F^{\,\mu\nu} F_{\mu\nu} + {\textstyle\frac12}\,\ve\,m^2\,A^\mu A_\mu
+ \ve\,\frac{b^\lambda\,b^\nu}{2b^2}\,F_{\lambda\rho} F_\nu^{\ \rho}\no
&-& {\alpha\over\pi}\;b_\mu\,A_\nu\,
\widetilde F^{\,\mu\nu} \qquad\quad \left(\ve={2\alpha b^2\over3\pi m^2}\right)\ .
\ea
\section{Summary of the Main Results}
To sum up, we would like to list the main achievements of our tough analysis
and make few more comments on consistency requirements and estimates for
the LIV vector components.

We recall that a temporal \AE ther $b^{\,\mu}$ is actually required
for a consistent fermion quantization \cite{Andrianov:2001zj,Adam:2001ma,Kostelecky:2000mm}.
However, in the lack of a photon mass and/or a bare CS vector of different direction,
it appears that a temporal \AE ther $b^{\,\mu}=(b,0,0,0)$
just leads \cite{Andrianov:1994qv,Andrianov:1998wj,Andrianov:1998ay,Adam:2001ma} to the instability of the photon
dynamics, that means imaginary energies for the soft photons.
 On the other hand, a spacelike vector $b^{\,\mu}$ causes problems for fermion quantization
\cite{Andrianov:2001zj,Kostelecky:2000mm} and thereby for the very meaning of the radiative
corrections.

To avoid this mismatch with classically massless photons
one can adopt the induced LIV
for {\sl lightlike} axial vectors $b^{\,\mu}\,.$ In particular,
for a lightlike universal axial vector
$b^{\,\mu}=(\,\pm\,b,{\bf b}\,)$ with $b=|{\bf b}|$ we find the dispersion relations
for the LIV 1-particle states of a fermion species $f$ that read
\begin{eqnarray}
p_+^0+b=\pm\sqrt{\Big({\bf p}+{\bf b}\,\Big)^2+m_f^2}\,,\qquad
p_-^0-b=\pm\sqrt{\Big({\bf p}-{\bf b}\,\Big)^2+m_f^2}\,.
\end{eqnarray}
Now, it turns out that the requirement $p_\pm^2>0$ for the LIV free 1-particle spinor physical states just drives to
the high momenta cutoff $|\,{\bf p}\,|\le m_e^2/4b\,.$
However, one has to keep in mind that the room for the existence of a universal
privileged spatial direction,
such as the \AE ther vector ${\bf b}\,,$
is essentially excluded by the torsion pendulum experiments with polarized electrons
\cite{Heckel:2006ww} which yield $|\,{\bf b}\,|\le5\times10^{-21}$ eV,
that should be compared to the benchmark value $m_e^2/M_{\rm Planck}=2\times10^{-17}$ eV.

Thus, the only left alternative to implement the LIV, with an essential contribution
from the fermion sector, consists in supplementing the photon dynamics
with a finite Lorentz invariant photon mass term so that
\be
m^2_{\,\gamma} =
\mu^2_{\,\gamma} + \Delta m^2_{\,\gamma} =
\mu^2_{\,\gamma} + \frac{2\alpha}{3\pi}\sum_f q_f^2\,b_f^2
\ee
for fermions with different charges $q_f$ and LIV four vectors $b_f\,,$
where $\mu_{\,\gamma}$ is a classical photon mass.
For a universal temporal vector $b^{\,\mu}_f\equiv b^{\,\mu}= (b,0,0,0)$
one finds the dispersion law
\ba
k_0^2 =
\left(\,|\,{\bf k}\,|\,\pm\,\frac{8\alpha}{\pi}\,b\,\right)^2
 + m^2_\gamma - b^2\,
\left\{\left(\frac{8\alpha}{\pi}\right)^2 +\
O\left(\frac{b\,|\,{\bf k}\,|}{m^2_e}\right)\right\}\,.
\ea
Hence, if $ m_\gamma \geq 8 \alpha\,b/\pi\,,$
then the photon energy keeps real for any wave vector $\bf k$  and LIV QED happens to be consistent
and the longitudinal polarization appears, due to the presence of a tiny photon mass $m_{\,\gamma}\,.$
Then the up to date very stringent experimental bound on the photon mass \cite{Amsler:2008zzb},
$m_\gamma < 6\times 10^{-17}$ eV, does produce the  limit $b < 3\times  10^{-15}\ {\rm eV}\,.$
Notice that in ref.~\cite{Kahniashvili:2008va} there is a claim that a closer bound on a
birefringent photon mass comes from the last five years results
on the oldest light of the universe, the Wilkinson Microwave Anisotropy Probe data,
{\rm viz.,} $m_{\;\!\gamma} < 3\times 10^{-19}$ eV, which entails the benchmark valued bound
$b< 2\times  10^{-17}$ eV and in turn $\Lambda_e\sim10^{28}\ {\rm eV}\sim M_{\rm Planck}\,.$

If the only source for LIV in QED is universally induced by fermions one can compare
these bounds with the following ones
which are required
to align the threshold of fermion UV instability to the existing experimental
data from LEP \cite{Hohensee:2009zk,Hohensee:2008xz}.
To fulfill it  one has to provide roughly  $ \mu\,m_e\,/\,2 b > 100$ GeV or $\mu > 4\times 10^{5}\,b\,.$
It gives a more stringent estimation $b < 10^{-22}\ {\rm eV}\,.$

Even the first bound entail the very large cutoff
$\Lambda_{\;\!e}\simeq {m_e^2/2b} \simeq 10^{26}$ eV whereas the second,
more stringent bound leads to the cutoff larger than the Planck mass scale.
It means that only the leading order in the LIV vector expansion makes any practical sense.

There are no better bounds on $b_\mu$ coming from the UHECRs data on the speed of light for photons.
This is because the increase of the speed of light depends quadratically on components of $b_\mu\,.$
Thus, for example, the data cited in \cite{Shore:2004sh,Jacobson:2005bg,Kostelecky:2008ts,Bietenholz:2008ni,Stecker:2009hj,Colladay:1996iz,
Colladay:1998fq,Kostelecky:2002hh,Shore:2003zc,Cheng:2006us} do imply much less severe
bounds on $|\,{\bf b}\,|$ or $b_0$ than those ones above mentioned.
In particular, the above reported field theoretical constraints for internal consistency
do actually lead to much more stringent limits on the temporal \AE ther $b^{\,\mu}=(b,0,0,0)$
than the most recent UHECRs data.

For the Lorentz invariance violating modification of QED, as discussed in the present paper,
the generic bounds
on LIV and CPT breaking parameters within the quantum gravity phenomenology
are less efficient to compete with the laboratory estimations.
They are dominating over other LIV effects in the high energy astrophysics.

An interesting bound on the deviations of the speed of light is given in \cite{Lieu:2003ee},
where spacetime fluctuations are addressed to produce modifications of the speed of light.
However, their estimation does not yield a better bound for a LIV vector $b_\mu\,,$
yielding, at best, $b_0 < 10^{-12}\ {\rm eV}\,,$
which is certainly less stringent as compared with the bounds discussed above.
\section{Conclusions and Final Discussion}
In this paper, the minimal model involving a Lorentz and CPT invariance violating
modification of quantum electrodynamics
has been thoroughly investigated. We have explicitly shown that the introduction
of a single LIV term $b_\mu\,\bar\psi(x)\,\gamma^{\,\mu}\,\gamma_5\,\psi(x)$
into the fermionic matter lagrangian does indeed give rise to a profound
modification of the whole theoretical model, in spite of the presently
allowed very small value for the temporal axial vector \AE ther background
$b_\mu=(b,0,0,0)$ with $b<3\times10^{-15}$ eV. Actually, the important lesson
we have learned from the present investigation
is that a tiny but non-vanishing breaking of the Lorentz and CPT symmetries does unavoidably
produce some drastic changes into the model, even of a nonperturbative nature.
As a paradigmatic example, it turns out that,
above the threshold $\Lambda_e\sim m_e^2/2b\,,$
electrons and positrons must decay by emitting tachyon like
photons and becoming left or right handed polarized. Since they are massive
their chiralities mix and therefore in a while they become again
vector like  and again decay. Thus, finally, high energy electrons and positrons
will be washed out from  the asymptotic Hilbert space and will not contribute
at all to the imaginary part of the vacuum polarization operator.
Of course, this might appear a little bit embarrassing,
as it hurts the long standing intuition and conventional wisdom,
as developed in the Lorentz invariant quantum field theory.
When one restores the full polarization operator with the help of the
dispersion relations, one should keep thereby only momenta lower than the threshold.

In this paper, which is essentially focussed
on the calculation of the 1-loop vacuum polarization tensor, we acknowledge that
a non-vanishing Chern-Simons vector and tiny photon mass are truly advocated and welcome
for : $i)$ the propagation of some LIV effect with a speed less than one is allowed;
$ii)$ the photodynamics is safely free from any acausal tachyonic effects;
$iii)$ the radiative corrections perfectly match with the existence of the
physical large momentum cutoff $\Lambda_e\,\lesssim\,M_{\rm Planck}\,,$ beyond which quantum gravity
might jeopardize the Einstein relativity principles;
$iv)$ the resulting effective quantum theory does fulfill locality, causality and unitarity
up to the physical large momentum cutoff $\Lambda_e$
that represents the intrinsic limit of the validity of the whole approach.
In conclusion, to the aim of seriously approaching any possible Lorentz invariance violation
for QED, one has to dismiss many of the well established
achievements as dictated by the application of the original Einstein critical
analysis about spacetime causality to the realm of the quantum field theory.

\section*{Acknowledgments}
This work was supported by Grants INFN IS-PI13 and INFN-MICINN Collaboration.
The work of A.A.A. was  supported by CUR Generalitat de Catalunya under project 2009SGR502,
by Grant FPA2007-66665,
by the Consolider-Ingenio 2010 Program CPAN (CSD2007-00042), and
by Grant RFBR 09-02-00073 and Program RNP2009-1575.
The work of J.A.  was partially supported by Fondecyt \# 1060646.
The work of M.C. was partially supported by projects DGAPA-UNAM-IN109107 and CONACYT \# 55310.
We are pleased to thank Dr. Fabrizio Sgrignuoli for his careful check of our calculations.
P.G. and R.S. wish to acknowledge the support of the Istituto Nazionale di Fisica Nucleare,
Iniziativa Specifica PI13, that contributed to the successful completion of this project.
\section*{Appendix A}
In this appendix we shall develop the technical details which are involved in
the evaluation of the 1-loop parity even part of the vacuum polarization tensor
in the constant temporal \AE ther $b_\mu\,.$
The trace in eq.~(\ref{5.1}) amounts to be
\begin{eqnarray}
&&\tr\left\{\bar\gamma^\mu\left(p^2+b^2-m^2
+ 2 b\cdot p\,\gamma_5
+ 2m b_\alpha\bar\gamma^\alpha\gamma_5\right)
\left(\gamma^\beta p_\beta
+ m
+ b_\beta\bar\gamma^\beta\gamma_5\right)\times\right.\no
&&\times\ \bar\gamma^\nu\left[\,(p+k)^2 + b^2 - m^2
+ 2 b\cdot (p+k)\gamma_5
+ 2m b_\lambda\bar\gamma^\lambda\gamma_5\,\right]\,\times\no
&&\left.\times\,\left[\,\gamma^\sigma(p+k)_\sigma
+ m
+ b_\sigma\bar\gamma^\sigma\gamma_5\,\right]\right\}\ ,
\label{5.2}
\end{eqnarray}
where we have taken into account that the external indices $\mu,\nu$
as well as the four-vector $b_\alpha$ are physical, \ie\
$\mu,\nu,\alpha=0,1,2,3$ so that, consequently, the corresponding
matrices $\bar\gamma^\mu,\bar\gamma^\nu$ are physical and contraction
of $b_\alpha$ with a $\gamma$ matrix always involves some
$\bar\gamma$-matrix, whereas
the loop momenta and any $\gamma$ matrix contracted with it
are $2\omega$ dimensional, \ie
$\p = p_\alpha \gamma^\alpha = \bar p_\alpha \bar \gamma^\alpha +
\hat p_\alpha \hat \gamma^\alpha$  -- see for instance~\cite{'tHooft:1972fi,Breitenlohner:1977hr,Breitenlohner:1975hg}.

The general structure of the photon self-energy tensor has been
already presented in eq.~(\ref{4.8}).
Here we are interested in the  Maxwell--Chern--Simons parity even term : the only
non-vanishing contributions to such a term are given by the traces of
the products of six, four and two gamma matrices, since the traces of the products of
three and five gamma matrices do indeed vanish
in $2\omega-$dimensions.
A straightforward computation gives, up to the overall factor $\tr\,{\mathbb I}=2^{\,\omega}\equiv4$
and keeping in mind that $\hat g^{\,\alpha\beta}p_\alpha\,p_\beta\,=\,-\,\hat p^2\,,$

\medskip\noindent
(a)\qquad six gamma matrices :
\begin{eqnarray}
&& +\ 4m^2\left\lbrace
b^2\left[\,p\cdot(p+k) + b^2\,\right]\ -\ 2b\cdot(p+k)\ b\cdot p\right\rbrace\,\bar g^{\,\mu\nu}\no
&& +\ 8m^2\,b\cdot k\ b^\nu\,p^\mu\
   +\ 8m^2\,b\cdot p\,\left(b^\mu\,k^\nu + b^\nu\,p^\mu + b^\mu\,p^\nu\right)\no
&& -\ 4m^2\,b^2\,\left(2p^\mu\,p^\nu + k^\mu\,p^\nu + k^\nu\,p^\mu\right)\
   -\ 8m^2\,(p^2 + p\cdot k)\,b^\mu\,b^\nu\no
&& -\ 8m^2\,\hat p^2\left(2b^\mu\,b^\nu\ -\ b^2\ \bar g^{\,\mu\nu}\right)\ ;
\label{5.3}
\end{eqnarray}

\medskip\noindent
(b)\qquad four gamma matrices :
\begin{eqnarray}
&& -\ 8\,b\cdot p\ b\cdot (p+k)\ \hat p^2\,\bar g^{\,\mu\nu}\no
&& +\ \left\{2p^\mu\,p^\nu\ +\ k^\mu\,p^\nu\ +\ k^\nu\,p^\mu\ -\
p\cdot(p+k)\ \bar g^{\,\mu\nu}\right\}\ \times\no
&& \times\ \left\lbrace
\left(p^2+b^2-m^2\right)\left[\,(p+k)^2 + b^2 - m^2\,\right]\
+\ 4\,b\cdot p\ b\cdot (p+k)\right\rbrace\no
&& +\ \left(2b^\mu\,b^\nu\ -\ b^2\,\bar g^{\,\mu\nu}\right)\left\lbrace
\left(p^2+b^2-m^2\right)\left[\,(p+k)^2 + b^2 - m^2\,\right]\right.\no
&& +\ 4\left(b\cdot p\right)\left[\,b\cdot(p+k)\,\right]\
+\ 2m^2\,\left.\left[\, p^2\ +\ (p+k)^2\ +\ 2b^2\,\right]\right\rbrace\no
&& +\  \left\lbrace 2\,b\cdot (p+k)\,\left(p^2+b^2-m^2\right) +
2\,b\cdot p\,\left[\,(p+k)^2 + b^2 - m^2\,\right]\right\rbrace\ \times\no
&& \times\ \left[\,2\,b\cdot p\ \bar g^{\,\mu\nu}\ -\
2\left(b^\mu\,p^\nu + b^\nu\,p^\mu\right)\
+\ b\cdot k\ \bar g^{\,\mu\nu}\ -\
b^\mu\,k^\nu -\ b^\nu\,k^\mu\,\right]\no
&& +\ 4m^2\,b\cdot p\,
\left[\,4\,b\cdot p\ \bar g^{\,\mu\nu}\ -\
2\left(b^\mu\,p^\nu + b^\nu\,p^\mu\right) -\
2b^\mu\,k^\nu\,\right]\no
&& +\  4m^2\,b\cdot k\,
\left[\,4\,b\cdot p\ \bar g^{\,\mu\nu}\ -\ 2b^\nu\,p^\mu\
+\ b\cdot k\ \bar g^{\,\mu\nu}\ -\ b^\mu\,k^\nu - b^\nu\,k^\mu\,\right]\no
&& -\ 2m^2\,b^2\,\left[\,p^2+(p+k)^2+2b^2-2m^2\,\right]\,\bar g^{\,\mu\nu}\ .
\label{5.5}
\end{eqnarray}

\medskip\noindent
(c)\qquad two gamma matrices :
\be
m^2\left\{\left(p^2+b^2-m^2\right)\left[\,(p+k)^2 + b^2 - m^2\,\right]
-\ 4\,b\cdot p\ \ b\cdot (p+k)\right\}\bar g^{\,\mu\nu}\ ;
\label{5.4}
\ee
Putting altogether we find
\ba
&& \reg \Pi_{\,{\rm even}}^{\ \mu\nu}(k,b,m,\mu)\ =\ 
4\,\i\,e^2 \mu^{4-2\omega}(2\pi)^{-2\omega}\ \times\no
&& \int\d^{\,2\omega}p\ {N^{\mu\nu}(p,k,b,m)}\,{D(p,b,m)\,D(p+k,b,m)}\ ,
\label{5.6}
\ea
where the scalar propagator reads
\begin{equation}
D(p,b,m)=\left\{(p^2+b^2-m^2+i\varepsilon)^2-4[\,(b\cdot p)^2-b^2
m^2\,]\right\}^{-1}\ ,
\label{5.10}
\end{equation}
so that
\ba
&&[\,D(p,b,m)\,D(p+k,b,m)\,]^{-1}\ =\no
&&\left\{(p^2+b^2-m^2+i\varepsilon)^2-4[\,(b\cdot p)^2-b^2m^2\,]\right\}\ \times\no
&&\left\{[\,(p+k)^2+b^2-m^2+i\varepsilon\,]^2-4[\,(b\cdot p+b\cdot k)^2-b^2m^2\,]\right\}\ .
\ea
Furthermore we get
\begin{eqnarray}
&& {N^{\mu\nu}(p,k,b,m)}\ =\no
&& -\ 8\,b\cdot p\ b\cdot (p+k)\ \hat p^2\,\bar g^{\,\mu\nu}\no
&& +\ \left\{2p^\mu\,p^\nu\ +\ k^\mu\,p^\nu\ +\ k^\nu\,p^\mu\ -\
\bar g^{\,\mu\nu}\ (p^2-m^2 + k\cdot p)\right\}\ \times\no
&& \times\ \left\lbrace
\left(p^2+b^2-m^2\right)\left[\,(p+k)^2 + b^2 - m^2\,\right]\
+\ \ 4\,b\cdot p\ b\cdot (p+k)\right\rbrace\no
&& +\ \left(2b^\mu\,b^\nu\ -\ b^2\,\bar g^{\,\mu\nu}\right)\left\lbrace
\left(p^2+b^2-m^2\right)\left[\,(p+k)^2 + b^2 - m^2\,\right]\right.\no
&& +\ 4\,b\cdot p\ b\cdot(p+k)\
+\ 2m^2\,\left.\left[\, p^2\ +\ (p+k)^2\ +\ 2b^2\
 -\ 4\hat p^2\,\right]\right\rbrace\no
&& +\  \left\lbrace 2\,b\cdot (p+k)\,\left(p^2+b^2-m^2\right) +
2\,b\cdot p\,\left[\,(p+k)^2 + b^2 - m^2\,\right]\right\rbrace\ \times\no
&& \times\ \left[\,2\,b\cdot p\ \bar g^{\,\mu\nu}\ -\
2\left(b^\mu\,p^\nu + b^\nu\,p^\mu\right)\
+\ b\cdot k\ \bar g^{\,\mu\nu}\ -\
b^\mu\,k^\nu -\ b^\nu\,k^\mu\,\right]\no
&& +\ 2m^2\,\bar g^{\,\mu\nu}\left[\, 2(b\cdot k)^2 - b^2(k^2-2m^2)\,\right]
-\  4m^2\,b\cdot k\,\left(\,b^\mu\,k^\nu +\ b^\nu\,k^\mu\right)\no
&& -\ 4m^2\,b^2\,\left(2p^\mu\,p^\nu + k^\mu\,p^\nu + k^\nu\,p^\mu\right)\
-\ 8m^2\,(\,p^2 + p\cdot k)\,b^\mu\,b^\nu\ .
\label{numeratore}
\end{eqnarray}

Now we can expand the polarization tensor in powers of the Lorentz symmetry
breaking axial vector $b_\lambda\,.$ To this aim, let us first define the following
quantities:
\ba
&& D(p,0,m)\ \equiv\ D(p,m)\ =\ (p^2-m^2+i\varepsilon)^{-2}\ ;\no
&& {N^{\mu\nu}(p,k,0,m)}\ \equiv\ {N_0^{\mu\nu}(p,k,m)}\no
&& =\ \left\{2p^\mu\,p^\nu\ +\ k^\mu\,p^\nu\ +\ k^\nu\,p^\mu\ -\
[\,p\cdot(p+k) - m^2\,]\ \bar g^{\,\mu\nu}\right\}\ \times\no
&& \times\ \left(p^2-m^2\right)\left[\,(p+k)^2 - m^2\,\right]\ ;
\ea
\ba
&& \frac{\partial}{\partial b_\lambda}\,D(p,b,m)\ =\no
&& -\ 4[\,D(p,b,m)\,]^2\left\lbrace b^\lambda\left(p^2+m^2+b^2\right)
-2p^\lambda\,b\cdot p\right\rbrace\ ;
\ea
\ba
&& \frac{\partial^2}{\partial b_\kappa\partial b_\lambda}\,D(p,b,m)\ =\
32\,[\,D(p,b,m)\,]^3\ \times\no
&&\times\ \left\lbrace b^\lambda\left(p^2+m^2+b^2\right) - 2p^\lambda\,b\cdot p\right\rbrace
\left\lbrace b^\kappa\left(p^2+m^2+b^2\right) - 2p^\kappa\,b\cdot p\right\rbrace\no
&& -\ 4[\,D(p,b,m)\,]^2\left\lbrace
g^{\kappa\lambda}\left(p^2+b^2+m^2\right) +2b^\lambda b^\kappa
-2p^\lambda p^\kappa\right\rbrace\ ;
\ea
\ba
&& D^{\prime\prime}(p,b,m)\ \equiv\ \frac12\,b_\kappa b_\lambda
\lim_{b\to 0}\,\frac{\partial^2}{\partial b_\kappa\partial b_\lambda}\,D(p;b,m)\no
&& =\ -\ 2[\,D(p,m)\,]^2\left\lbrace
b^2\left(p^2 + m^2\right) - 2 (b\cdot p)^2\right\rbrace\ ;
\ea
\ba
&& {N_2^{\mu\nu}(p,k,b,m)}\ \equiv\ \frac12\,b_\kappa b_\lambda
\lim_{b\to 0}\,{\partial^2\over \partial b_\kappa\partial b_\lambda}\,
{N^{\mu\nu}(p,k,b,m)}\no
&& =\ -\ 8\,b\cdot p\ b\cdot (p+k)\ \hat p^2\,\bar g^{\,\mu\nu}\no
&& +\ \left\{2p^\mu\,p^\nu\ +\ k^\mu\,p^\nu\ +\ k^\nu\,p^\mu\ -\
\bar g^{\,\mu\nu}\ (p^2-m^2 + k\cdot p)\right\}\ \times\no
&& \times\ \left\lbrace b^2\left[\,(p+k)^2 + p^2 - 2m^2\,\right]\
+\ \ 4\,b\cdot p\ b\cdot (p+k)\right\rbrace\no
&& +\ \left(2b^\mu\,b^\nu\ -\ b^2\,\bar g^{\,\mu\nu}\right)\ \times\no
&& \times\ \left\lbrace
\left(p^2-m^2\right)\left[\,(p+k)^2 - m^2\,\right]
+\ 2m^2\left[\, p^2\ +\ (p+k)^2\ -\ 4\hat p^2\,\right]\right\rbrace\no
&& +\  \left\lbrace 2\,b\cdot (p+k)\,\left(p^2-m^2\right) +
2\,b\cdot p\,\left[\,(p+k)^2 - m^2\,\right]\right\rbrace\ \times\no
&& \times\ \left[\,2\,b\cdot p\ \bar g^{\,\mu\nu}\ -\
2\left(b^\mu\,p^\nu + b^\nu\,p^\mu\right)\
+\ b\cdot k\ \bar g^{\,\mu\nu}\ -\
b^\mu\,k^\nu -\ b^\nu\,k^\mu\,\right]\no
&& +\ 2m^2\,\bar g^{\,\mu\nu}\left[\, 2(b\cdot k)^2 - b^2(k^2-2m^2)\,\right]
-\  4m^2\,b\cdot k\,\left(\ b^\mu\,k^\nu +\ b^\nu\,k^\mu\right)\no
&& -\ 4m^2\,b^2\,\left(2\,p^{\,\mu}\,p^\nu + k^\mu\,p^\nu + k^\nu\,p^\mu\right)\
-\ 8m^2\,(p^2 + p\cdot k)\,b^\mu\,b^\nu\ .
\label{definitions}
\ea

It follows that we can write, up to the lowest order
approximation,
\ba
&& \reg \Pi_{\,{\rm even}}^{\,\mu\nu}(k,b,m,\mu)\ =\no
&& \reg \left\{\Pi_{\,{\rm cov}}^{\,\mu\nu}(k,m,\mu)\ +\
   \Delta \Pi_{\,{\rm even}}^{\,\mu\nu}(k,b,m,\mu)\right\}\
+\ {\rm O}(b^2)^2\ ,
\ea
in which we have set
\ba
\reg \Pi_{\,{\rm cov}}^{\,\mu\nu}(k,m,\mu)&=& 
4\,\i\,e^2\int_p
{N_0^{\mu\nu}(p,k,m)}\,{D(p,m)\,D(p+k,m)}\no
&=&\left(k^2 g^{\,\mu\nu} - k^\mu k^\nu\right)\,\reg \Pi(k,m)\ ,
\label{covapola}
\ea
where
\be
\int_p\ \equiv\ \mu^{\,2\epsilon}\;(2\pi)^{-2\omega}\int {\rm d}^{2\omega}p\ ,\qquad\quad
\epsilon = 2-\omega\ .
\ee
As it is well known -- see \eg \cite{Peskin:1995ev} -- the gauge invariant
part of the dimensionally regularized Lorentz invariant polarization function is provided by
\ba
\reg \Pi(k,m)&=& -\;\frac{8e^2\mu^{\,2\epsilon}}
{(4\pi)^\omega}\int_0^1{\rm d}x\
{x(1-x)\,\Gamma(2-\omega)}{\left[\,m^2-x(1-x)k^2\,\right]^{\omega-2}}\no
&=& -\ {\alpha\over 3\pi}\, \left\lbrace\frac{1}{\epsilon} - {\bf C} + \ln{4\pi\mu^2\over m^2}
-\int_0^1{{\rm d}x\over R}\left(4x^4-8x^3+3x^2\right)\right\rbrace\no
&+& {\rm irrelevant\ for\ } \epsilon\to 0\ ,
\ea
with\quad $R=x^2-x+m^2/k^2\,.$ Notice that we have
\ba \int_0^1{{\rm d}x\over R}\left(4x^4-8x^3+3x^2\right) \buildrel k\to 0
\over \sim\ -\ {k^2\over 5m^2}
\ea
so that the well known Uehling-Serber result is recovered
\be
\reg \Pi(k,m)- \reg \Pi(0;m^2)\ =\ {-\,\alpha k^2\over 15\pi m^2}\ +\ O\left({k^2\over m^2}\right)^2
\ee
The covariant divergent part can be removed from vacuum polarization by adding the usual local counterterm
\be
{\cal L}_{\,\rm c.t.}=-\,\frac14\,(Z_3-1)\,F^{\,\mu\nu}F_{\mu\nu}
\ee
where the lowest order photon wave function renormalization constant
in the BPHZ on the mass shell subtraction scheme \cite{Pokorski:1987ed} is given by
\be
Z_3 = 1-{\alpha\over 3\pi}\, \left({1\over \epsilon}-{\bf C}+\ln{4\pi\mu^2\over m^2}\right)+O(\epsilon)
\ee
whence we can immediately segregate the divergent part of the
${\rm  O}(\alpha)$ shift in the electric charge \cite{Peskin:1995ev}
\be
Z_3-1=\reg \Pi(0;m^2)+{\rm O}(\alpha^2)\;\approx\;
{-\,\alpha\over 3\pi\epsilon}\ ,
\ee
so that  we eventually come to the well known -- see \eg
refs.~\cite{Pokorski:1987ed,Peskin:1995ev} --
finite expression for the polarization invariant function in the BPHZ
on the mass shell subtraction scheme
\ba \widehat\Pi(k,m)&=& {\alpha\over 3\pi}\, \left\lbrace\frac43
-\int_0^1{{\rm d}x\over R}
\left[\,4x^3-x^2\left(3-{4m^2\over k^2}\right)\,\right]\right\rbrace\ .
\ea

The  first correction to the even part of the vacuum polarization tensor
(formally) violating Lorentz and CPT invariance reads
\ba
\reg\ \Delta \Pi_{\,{\rm even}}^{\,\mu\nu}(k,b,m,\mu) &=&
4\,\i\,e^2\mu^{\,2\epsilon}\int{\d^{2\omega}p\,(2\pi)^{-2\omega}}\ \times\no
&\times& \left\lbrace N_0^{\mu\nu}(p;k,m)\,D^{\prime\prime}(p;b,m)\,D(p+k;m)\right.\no
&+& N_0^{\mu\nu}(p;k,m)\,D(p;m)\,D^{\prime\prime}(p+k;b,m)\no
&+& \left.N_2^{\mu\nu}(p;k,b,m)\,D(p;m)\,D(p+k;m)\right\rbrace\no
&\equiv& 2I^{\mu\nu} + J^{\mu\nu}
\ea
According to eq.~(\ref{definitions}) we obtain
\ba I_{\,\mu\nu} &=& -\,8\i\,e^2\int_p\
\left\{2p^\mu\,p^\nu + k^\mu\,p^\nu + k^\nu\,p^\mu- [\,p\cdot(p+k) - m^2\,]\
\bar g^{\,\mu\nu}\right\}\no
&\times& \frac{b^2\left(p^2 - m^2\right) - 2 (b\cdot p)^2 + 2b^2m^2}
{\left(p^2 - m^2 + i\varepsilon\right)^3
\left[\,(p+k)^2 - m^2 + i\varepsilon\,\right]}\ ;
\label{I1}
\ea
\ba
J^{\mu\nu}(k;b,m) &=& \int_p\ \frac{-\,16\,\i\,e^2}{\left(p^2 - m^2 + i\varepsilon\right)^2
\left[\,(p+k)^2 - m^2 + i\varepsilon\,\right]^2}\no
&\times& \left\lgroup 2\,b\cdot p\ b\cdot (p+k)\ \hat p^2\,\bar g^{\,\mu\nu}\right.\no
&-& {\textstyle\frac14}\,\left\{2p^\mu\,p^\nu\ +\ k^\mu\,p^\nu\ +\ k^\nu\,p^\mu\ -\ \bar g^{\,\mu\nu}\,
(p^2-m^2 + p\cdot k)\right\}\ \times\no
&\times&  \left\lbrace b^2\left[\,(p+k)^2 + p^2 - 2m^2\,\right]\ +\
4\,b\cdot p\ b\cdot (p+k)\right\rbrace\no
&-& {\textstyle\frac14}\,\left(2b^\mu\,b^\nu\ -\ b^2\,\bar g^{\,\mu\nu}\right)\ \times\no
&\times& \left\lbrace \left(p^2-m^2\right)\left[\,(p+k)^2 - m^2\,\right] +\
2m^2\left[\, p^2\ +\ (p+k)^2\ -\ 4\hat p^2\,\right]\right\rbrace\no
&-& {\textstyle\frac12}\,\left\lbrace  b\cdot (p+k)\,\left(p^2-m^2\right) + b\cdot p\,
\left[\,(p+k)^2 - m^2\,\right]\right\rbrace\ \times\no
&\times& \left[\,2\,b\cdot p\ \bar g^{\,\mu\nu}\ -\ 2\left(b^\mu\,p^\nu + b^\nu\,p^\mu\right)\
+\ b\cdot k\ \bar g^{\,\mu\nu}\ -\ b^\mu\,k^\nu -\ b^\nu\,k^\mu\,\right]\no
&-&  {\textstyle\frac12}\,m^2\,\bar g^{\,\mu\nu}\left[\,2(b\cdot k)^2-b^2(k^2-2m^2)\,\right]
+ m^2\,b\cdot k\, \left(b^\mu\,k^\nu +\ b^\nu\,k^\mu\right)\no
&+& m^2\,b^2\,\left(2p^\mu\,p^\nu + k^\mu\,p^\nu + k^\nu\,p^\mu\right)\ +\
2m^2\,(p^2 + p\cdot k)\,b^\mu\,b^\nu\left.\right\rgroup\ .
\label{I3}
\ea
\section*{Appendix B}
In this appendix we aim to enter in the details of the calculation that leads to the
appearance of the induced photon mass at one loop using the so called $\overline{DR}$
version of dimensional regularization.
Actually, it turns out that a possible non-vanishing photon mass might be radiatively generated to the lowest order,
within the LIV modification of QED and using dimensional regularization
to deal with divergencies, if and only if
\ba
\reg\ \Delta \Pi_{\,{\rm even}}^{\,\mu\nu}(k=0,b,m,\mu)\equiv
\reg\ \Delta \Pi_{\,{\rm even}}^{\,\mu\nu}(b,m,\mu)\not=0\ .
\label{LSBmass}
\ea

Now we have
\ba I^{\mu\nu}(0;b,m) &\equiv& I_0^{\mu\nu}(b,m) = -\,8\,\i\,e^2\int_p\ \left[\,2p^\mu\,p^\nu - (p^2 - m^2)\ \bar g^{\,\mu\nu}\,\right]\no &\times& \frac{b^2\left(p^2 - m^2\right) - 2 (b\cdot p)^2 + 2b^2m^2}{\left(p^2 - m^2 + \i\varepsilon\right)^4}\ . \ea
so that can we rewrite the above expression in terms of the basic integrals: namely,
\ba
I_0^{\mu\nu}(b,m) &=& -\,16\,\i\,e^2\Big\lbrace b^2\,I^{\mu\nu}(3,0)-2b_\rho b_\sigma\,I^{\mu\nu\rho\sigma}(4,0) + 2b^2m^2\,I^{\mu\nu}(4,0)\no
&-& {\textstyle\frac12}\,\bar g^{\,\mu\nu}b^2\,I(2,0) + \bar g^{\,\mu\nu}b_\rho b_\sigma\,I^{\rho\sigma}(3,0)
- b^2m^2\bar g^{\,\mu\nu}\,I(3,0)\Big\rbrace\ ,
\ea
where the basic covariant integral of generic tensor rank is defined by
\be
I^{\mu\nu\,\cdots\,\rho\sigma}(n,0)\equiv
\int_p\frac{p^{\,\mu}\,p^{\,\nu}\,\cdots\,p^{\,\rho}\,p^{\,\sigma}}{\left(\,p^2 - m^2 + \i\varepsilon\,\right)^n}
\ee
In a quite similar way we obtain
\ba
&& J^{\mu\nu}(0;b,m) \equiv J_0^{\mu\nu}(b,m) =
{4\,\i\,e^2}\, \int_p\ {\left(p^2 - m^2 + \i\varepsilon\right)^{-4}}\no
&\times& \left\lbrace -\ 8\,(b\cdot p)^2\ \hat p^2\,\bar g^{\,\mu\nu}\ \right. +\
m^2\,\bar g^{\,\mu\nu}\,\left[\,2b^2\left(p^2 - m^2\right) +\ 4\,(b\cdot p)^2\,\right]\no
&+& \left(2p^\mu\,p^\nu\ - p^2\ \bar g^{\,\mu\nu}\right)
\left[\,2b^2\left(p^2 - m^2\right)\ +\ 4\,(b\cdot p)^2\,\right]\no
&+& \left(2b^\mu\,b^\nu\ -\ b^2\,\bar g^{\,\mu\nu}\right) \left[\,\left(p^2-m^2\right)^2 +\
2m^2\left(2p^2\ -\ 4\hat p^2\right)\,\right]\no
&+& 8\,b\cdot p\,\left(p^2-m^2\right) \left(b\cdot p\
\bar g^{\,\mu\nu}\ -\ b^\mu\,p^\nu\ -\ b^\nu\,p^\mu\right)\no
&+& 4\,b^2 m^4\,\bar g^{\,\mu\nu}\ -\ 8m^2\,b^2\,p^\mu\,p^\nu\ -\ \left. 8m^2\,p^2\,b^\mu\,b^\nu\right\rbrace
\ea
and in terms of the basic integrals we can eventually write
\ba
&& J_0^{\mu\nu}(b,m)\ =\ 16\,\i\,e^2\,\bar g^{\,\mu\nu}\ \times\no
&\times& \Big\lbrace 2b_\rho b_\sigma\,\hat g_{\lambda\kappa}\,I^{\lambda\kappa\rho\sigma}(4,0) -
{\textstyle\frac34}\,b^2\,I(2,0) + b_\rho b_\sigma\,I^{\rho\sigma}(3,0) - b^2m^2I(3,0)\Big\rbrace\no
&-& 16i\,e^2\Big\lbrace - b^2\,I^{\mu\nu}(3,0) - 2b_\rho b_\sigma\,I^{\mu\nu\rho\sigma}(4,0) +
2b^2\,m^2\,I^{\mu\nu}(4,0)\no
&-& {\textstyle\frac12}\,b^\mu b^\nu\,I(2,0) + 2b^\mu b_\rho\,I^{\nu\rho}(3,0)
+ 2b^\nu b_\rho\,I^{\mu\rho}(3,0)\Big\rbrace\no
&+& {\rm evanescent\ terms\ for\ \omega\to 2}\ .
\ea
To sum up, the lowest order Lorentz non-invariant mass term can be reduced
to the following combination of the basic integrals: namely,
\ba
&& \reg\ \Delta \Pi_{\,{\rm even}}^{\,\mu\nu}(b,m,\mu) = 2I_0^{\mu\nu}(b,m)+J_0^{\mu\nu}(b,m)
= 16\,\i\,e^2\,\bar g^{\,\mu\nu}\ \times\no
&\times& \Big\lbrace 2b_\rho b_\sigma\,\hat g_{\lambda\kappa}\,I^{\lambda\kappa\rho\sigma}(4,0)
+ {\textstyle\frac14}\,b^2\,I(2,0) - b_\rho b_\sigma\,I^{\rho\sigma}(3,0) + b^2m^2I(3,0)\Big\rbrace\no
&-& 16\,\i\,e^2\Big\lbrace b^2\,I^{\mu\nu}(3,0) - 6b_\rho b_\sigma\,I^{\mu\nu\rho\sigma}(4,0)
+ 6b^2\,m^2\,I^{\mu\nu}(4,0)\no
&-&{\textstyle\frac12}\,b^\mu b^\nu\,I(2,0) + 2b^\mu b_\rho\,I^{\nu\rho}(3,0)
+ 2b^\nu b_\rho\,I^{\mu\rho}(3,0)\Big\rbrace\no
&+&{\rm evanescent\ terms\ for\ \omega\to 2}\ .
\label{dimphotonmass}
\ea

\end{document}